\documentclass[fleqn,usenatbib]{mnras}

\usepackage{newtxtext,newtxmath}

\usepackage[T1]{fontenc}
\usepackage{ae,aecompl}
\usepackage{bm,dsfont}
\usepackage{hyperref}
\usepackage{url}
\usepackage{subfig}
\usepackage{graphicx}

\newcount\colveccount
\newcommand*\colvec[1]{
        \global\colveccount#1
        \begin{pmatrix}
        \colvecnext
}
\def\colvecnext#1{
        #1
        \global\advance\colveccount-1
        \ifnum\colveccount>0
                \\
                \expandafter\colvecnext
        \else
                \end{pmatrix}
        \fi
}

\newcommand{\Tr}{\mathrm{Tr}}

\usepackage{amsmath}	
\usepackage{amssymb}	

\title[Mapping Incoherent GWBs]{Mapping Incoherent Gravitational Wave Backgrounds}

\author[A. I. Renzini et al.]{
A. I. Renzini,\thanks{E-mail: arianna.renzini15@imperial.ac.uk}
and C. R. Contaldi
\\
$^{1}$Blackett Laboratory, Imperial College London, South Kensington Campus, London, SW7 2AZ, UK
}

\pubyear{2018}

\begin{document}
\label{firstpage}
\pagerange{\pageref{firstpage}--\pageref{lastpage}}
\maketitle

\begin{abstract}
Given the recent detection of gravitational waves from individual sources it is almost a certainty that some form of background of gravitational waves will be detected in future. The most promising candidate for such a detection are  backgrounds made up of incoherent superposition of the signal of unresolved astrophysical or, backgrounds sourced by earlier cosmological events. Such backgrounds will also contain anisotropies about an average value. The information contained in the background level and any anisotropies will be extremely valuable as an astrophysical and cosmological probe. As such, the ability to reconstruct sky maps of the signal will become important as the sensitivity increases. We build and test a pixel--based, maximum--likelihood Gravitational Wave Background (GWB) map-maker that uses the cross-correlation of sets of generalised baselines as input. The resulting maps are a representation of the GWB power, or strain ``intensity'' on the sky. We test the algorithm by reconstructing known input maps with different baseline configurations. We also apply the map-maker to a subset of the Advance LIGO data.
\end{abstract}


\section{Introduction}\label{sec:Intro}

There have now been a number of confirmed, direct detections of gravitational wave signals \citep{Abbott2016a,Abbott2016b,Abbott2016c,Abbott2017e,Abbott2017a,Abbott2017b,Abbott2017c}. The search for other events is continuing with detectors at three locations; the Laser Interferometer Gravitational Wave observatory (LIGO) detectors \citep{TheLIGOScientific:2016agk} in Hanford (WA) and Livingston (LA), and the Virgo detector in Pisa, Italy \citep{2011CQGra..28k4002A}. So far, the signals detected have been interpreted as transient signals emitted during the final moments of compact object mergers. Such events are identified via signal-to-noise threshold triggers and then classified through comparisons against predicted signal templates. As such they represent a tiny fraction of the data collected by the network of detectors.  Whilst it is very difficult to identify signals of {\sl individual} events below the detection threshold, given enough sensitivity and integration, it is expected that these events will eventually lead to the statistical detection of a background made up of the superposition of {\sl many} undetected events. Indeed, there are a number of astrophysical sources that should contribute to such a background \citep[see e.g.][]{Regimbau2011}. Cosmological and primordial backgrounds may also be present, albeit at levels far below astrophysical ones at accessible frequencies in the foreseeable future \citep[see e.g.][]{Caprini2018}.

The detection of an astrophysical background will be challenging with LIGO class detectors although the number of such detectors is expected to double over the next decade with the addition of the Kamioka Gravitational Wave Detector (KAGRA) in the Kamioka mine, Japan \citep{Aso:2013eba}, the GEO600 detector in Sarstedt, Germany \citep{Dooley:2015fpa}, and the Indian Initiative in Gravitational-wave Observations (IndIGO) \citep{Unnikrishnan:2013qwa}. This, together with the guarantee of longer integration times sustained by the increase in funding post-detection, may bring the overall, integrated sensitivity to a point where a background detection may be feasible. Irrespective of this, pulsar timing arrays (PTAs) are expected to yield direct detection of a GWB in the nano-hertz frequency band over the next 10 years \cite{Taylor2016}. PTAs are particularly relevant in gravitational wave astronomy as direct detection of gravitational wave signatures is possible by analysing the timing residuals of millisecond pulsars \citep[see e.g.][]{Hobbs2009}. Specifically, PTAs are well suited to characterising a GWB of stochastic origin \citep{Mingarelli2013}.  On longer time scales, the Laser Interferometer Space Antenna (LISA) mission \citep{Cruz2005,Ricciardone2017}, with an expected launch date in the late 2030s, is virtually guaranteed to reach the  sensitivity required to detect a background of galactic and extra galactic sources. In fact, the presence of these backgrounds constitutes a challenge for the LISA  analysis pipeline as the presence of a statistical background complicates the problem of signal and noise estimation. These problems are similar to ones encountered in Cosmic Microwave Background (CMB) data analysis \citep[see e.g.][]{Adam:2015rua} and the analogy is particularly close when considering the case of low frequency interferometric measurements of the CMB using coherent detectors in radio interferometry
\citep[see e.g.][]{1999ApJ...514...12W,Halverson:2001yy,Myers:2002tn}. 

A detection and characterisation of both stochastic and primordial GWBs is an exciting prospect. The average value of stochastic backgrounds, together with any spectral dependence, will yield important constraints on the redshift distribution and nature of the source population. Beyond this, any detection of anisotropies in the backgrounds, particularly for extra--galactic sources, will constitute a biased tracer of the underlying matter distribution and will open a new, multi--messenger, window on studies of large scale structure. In the case of primordial backgrounds, the detection of an average gravitational wave density would constitute a ground--breaking result and any information on anisotropies could yield direct constraints on the Planck epoch. Efforts to calculate the expected anisotropies in a variety of GWBs are ongoing \citep{Contaldi:2016koz,Cusin:2017fwz}.

In light of this and the expected growth in detector numbers and sensitivity we have initiated a program aimed at building a generalised, maximum--likelihood, gravitational wave map--making algorithm for the purpose of mapping {\sl incoherent} backgrounds. Our program uses a bottom-up approach to dealing with data and works directly in the sky--pixel domain building on the experience of similar efforts in CMB analysis. 

Algorithms to map the gravitational wave density and its directional dependence onto the sky frame have been considered by a number of authors in the literature including \cite{Christensen1992,Cornish2001,Ballmer2006,Mitra2008,Thrane2009,Gair:2014rwa,Romano2015,Romano2017}. Actual constraints on the background energy density of gravitational waves $\Omega_{\rm GW}$ and its anisotropies have been published by the LIGO collaboration in \cite{Abbott2009,Abbott2017d,Abbott2017c}. These have employed a number of methods such as sidereal folding of the data \citep[see also][]{Ain2015}, and radiometer or spherical harmonic domain methods \cite{Thrane2009} using the frame imposed by the LIGO baseline to reduce the sky rotation to a single phase modulation. More generalised methods have been explored by \cite{Cornish2001,Romano2017}. These have included phase--coherent methods of \cite{Romano2015} and a well--developed ``radiometer'' method aimed at mapping the sky signal of individually occurring sources along with algorithms that constrain directly the spherical harmonic coefficients of the GWB \citep{Ballmer2006,Thrane2009}. 

The approach described here is distinct from these although most similar in scope and aim to that of \cite{Mitra2008} and \cite{Thrane2009}. Our algorithm will reconstruct maximum--likelihood maps of GWB power directly in the sky--oriented, pixel frame, using the cross-correlation of the signals of generalised detectors. The maps produced will be of the ``intensity'' \footnote{We introduce this in analogy to the electro--magnetic intensity as the second order ensemble average of the underlying  field amplitude.} of the strain signal, {\sl i.e.} second order in the strain $h$, as a function of direction on the sky. The method is therefore phase--{\sl incoherent} since it discards the phase information present in the individual, coherent detector measurements. It is similar in nature to the use of coherent measurements in interferometry to map the intensity of the CMB.   

The outline of this {\sl paper} is as follows; in Section~\ref{sec:GWB} we review the nature and fundamental properties of the gravitational wave signal we are targeting together with the formalism describing the observations using cross--correlation of the measurements of generalised gravitational wave detectors. In Section~\ref{sec:MaMa} we describe the maximum--likelihood map reconstruction procedure and adapt it specifically to gravitational wave measurements as discussed in Section~\ref{sec:GWB}. In Section~\ref{sec:res} we test our algorithm on simulated data modelled on existing LIGO data. We do this for a number of simulated signals to test the reconstruction of different types of structures on the sky. We also run the algorithm on a small subset of the available LIGO data. We conclude with a short discussion of our results and a summary of ongoing work in Section~\ref{sec:disc}.

\section{The signal of Gravitational Wave Backgrounds}\label{sec:GWB}

GWBs are usually characterised by the dimensionless energy density $\Omega_{\rm GW}$  \citep[see e.g.][]{Allen1999}. The spectral dependence of this measure is given by the physical energy density of the GWB per logarithmic frequency interval
\begin{equation}
\Omega_{\rm GW}(f) = \frac{1}{\rho_c} \frac{d\rho_{\rm GW}}{d\ln f}\,,
\end{equation}
$\rho_c$ is the critical energy density. 
Beyond the isotropic value of the GWB over the entire sky we can include any anisotropy by adding a directional dependence $\Omega_{\rm GW}(f) \rightarrow \Omega_{\rm GW}(f,\bm{\hat n})$ where $\bm{\hat n}$ is the unit vector of a line-of-sight on the sky.

GWBs fall broadly into two distinct types based on the underlying generation mechanism. The first is a superposition of the signal of astrophysical sources, both galactic and extra--galactic, that are not individually detected or resolved. This is also known as a {\sl stochastic} GWB component. This background is directly linked to the source event rate as a function of redshift, as well as the spectral energy density of the source population \citep{Regimbau2011}. 

The second category is a cosmological or primordial GWB. This category covers GWBs generated by the evolution of vacuum fluctuations during an inflationary epoch that end up as super--horizon tensor modes at the end of inflation \citep{Grischuk1975,Maggiore1999}. It also includes GWBs generated by non-linear phenomena at early times such as phase transitions or via emission by topological defects \citep{Hogan1986,Battye1997,Vilenkin2000}.

Whilst GWBs of different origin will have differing spectral dependence, most will be unpolarised and incoherent in that the temporal phase of the signal is expected to be random. A possible exception for this is a truly primordial GWB i.e. one that is formed by modes that at one time were frozen--out on super--horizon scales. In this case the GWB should form standing waves and the phase is correlated between modes with opposite momenta \citep{Contaldi2018}. 

Throughout this {\sl paper} we will assume the GWB signal being mapped has no frequency dependence. This assumption renders the effective weighting of our estimation process sub--optimal for any other spectral dependence. In a realistic search for a GWB signal the map-making process would be repeated using various assumptions for the underlying spectral dependence or, given enough spectral resolution, the signal could be binned into separate frequency bands to produce a set of maps. Here we will integrate all signals across a single range in frequency. 

To test our procedure we will use injected signals from different input maps in order to verify the reconstruction of a known sky. One form of input will be from stationary maps of strain intensity of varying amplitudes and with anisotropies given by a Gaussian random field with a scale invariant power spectrum. These are not to be regarded as realistic simulations of any particular GWB component and are used solely for testing purposes. We also use non-stationary maps where the signal is made up of transient events drawn from Poisson distributions whose mean is given by an underlying anisotropic map. These are also not supposed to be realistic simulations of a GWB but show how such a signal could arise from an actual process that is stochastic in the time domain. 

\subsection{Strain signal}\label{subsec:Pol}

The transverse, traceless strain tensor $h_{ij}$ at time $t$ and position vector $\bm x$ can be decomposed into two independent, spin-2, polarisation states  $h_+$ and $\,h_{\times}$ and expanded using plane waves as
\begin{equation}
h_{ij}\,(t,\bm{x})=\int_{-\infty}^{+\infty} \!\!\!df \int_{S^2} \!\!\!d\bm{\hat n}\!\!\sum_{P=+,\,\times}\!\!h_P\,(f,\,\bm{\hat n})\,e_{ij}^P(\bm{\hat n})\,e^{i2\pi f(\bm{n\cdot x}-t)}\,,
\label{expan}
\end{equation}
where polarisation base tensors $e^P$ are
\begin{align}
e^+ =e_\theta\otimes e_\theta - e_\phi\otimes e_\phi \,,&&\\
e^\times = e_\theta\otimes e_\phi+e_\phi\otimes e_\theta\,,&&
\end{align}
with
\begin{equation}
\begin{split}
e_\theta &= (\cos\theta\cos\phi,\cos\theta\sin\phi,-\sin\theta) \,,\\
e_\phi &= (-\sin\phi,\cos\phi,0)\,,
\end{split}
\end{equation}
and we are working in units where $c=1$.

For a stochastic or cosmological GWB $h_P$ represents a random amplitude. We will assume the amplitudes are drawn from a Gaussian probability distribution. This is most likely a good approximation for a cosmological GWB. For a stochastic GWB of astrophysical sources this assumption may break down but the central limit theorem will guarantee that the statistics approach that of a Gaussian random field if the signal is sourced by a sufficiently large number of independent events and that any high signal-to-noise outliers have been subtracted from the detector time streams. Under the Gaussian assumption the statistical properties of the amplitudes are then characterised solely by the second order moments\footnote{In the limit that the signal is non-Gaussian the maximum--likelihood maps derived below remain unchanged but the interpretation of their covariance would be affected by the presence of higher order cumulants in the underlying probability densities.} $\langle h_P^{}(f,\,\bm{\hat n}) h'^\star_{P'}(f',\,\bm{\hat n}')\rangle$, which correspond to ensemble averages
\begin{equation}
\begin{split}
 \begin{pmatrix} \langle h^{}_+\,h'^\star_+\rangle  & \langle h^{}_+\,h'^\star_\times\rangle  \\ \langle h^{}_\times\,h'^\star_+\rangle  & \langle h^{}_\times\,h'^\star_\times\rangle  \end{pmatrix} &= \frac{1}{2}\,\delta(\bm{n-n'})\,\delta(f-f')\,\times\\
 &\begin{pmatrix} I(f,\,\bm{\hat n})+Q(f,\,\bm{\hat n}) & U(f,\,\bm{\hat n})-iV(f,\,\bm{\hat n}) \\ U(f,\,\bm{\hat n})+iV(f,\,\bm{\hat n}) & I(f,\,\bm{\hat n})-Q(f,\,\bm{\hat n}) \end{pmatrix}\,,
\end{split}
\label{stokey}
\end{equation}
where we have introduced the Stokes parameters $I$, the intensity, $Q$ and $U$, giving the linear polarisation, and $V$, the circular polarisation. The four Stokes parameters completely describe the polarisation of the observed signal in analogy with electromagnetic Stokes parameters for the photon. The difference here is that whilst the electromagnetic $Q$ and $U$ Stokes parameters transform as spin-2 quantities with respect to rotations, their strain counterparts transform as spin-4 under rotations. In both cases the intensity $I$ behaves as a scalar under rotations.

\begin{figure}
\centering
\includegraphics[width=\columnwidth]{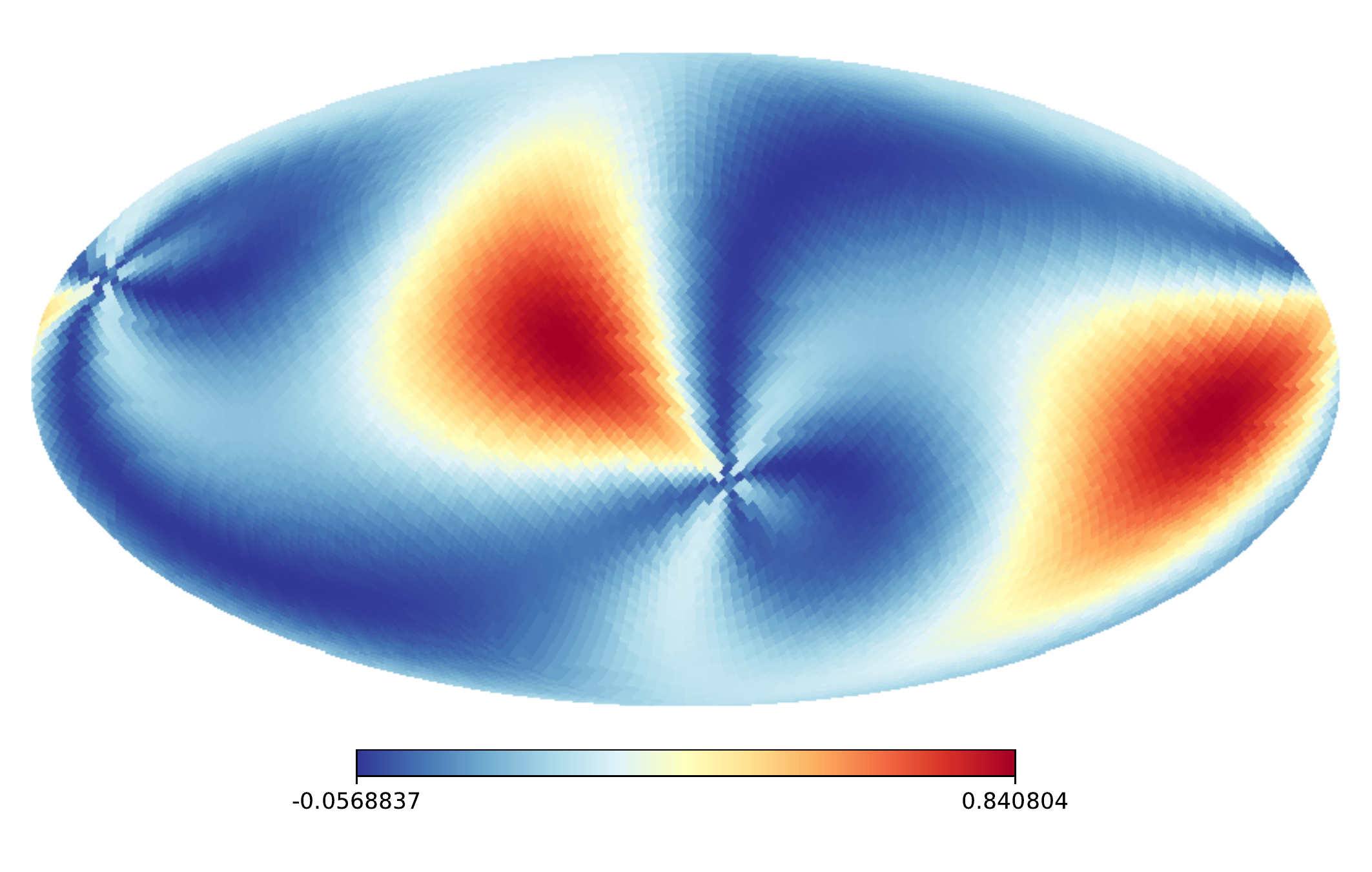}
\includegraphics[width=\columnwidth]{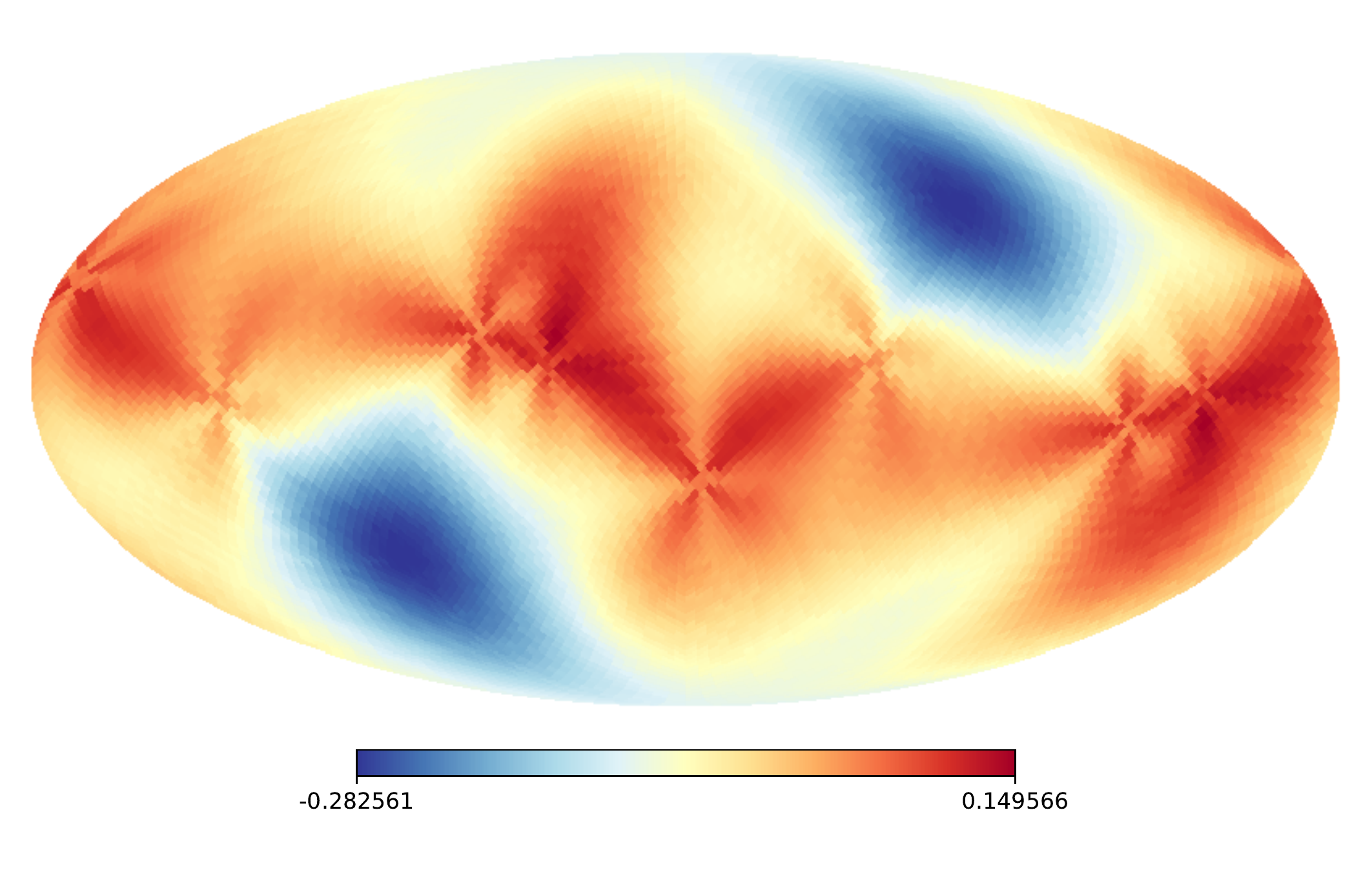}
\caption{Overlap function $\gamma(\bm\hat n)$ for the single LIGO baseline ({\sl top}) and the sum of overlap functions for the ten baselines of the five detector combination listed in Table~\ref{tab:dects} ({\sl bottom}). Both maps are in galactic coordinates and show how the strain intensity is integrated at an arbitrary phase of the Earth's rotation.}
\label{fig:overlap}
\end{figure}

In the following we will restrict our analysis to the reconstruction of the GWB intensity $I(f,\,\bm{\hat n})$ only, corresponding to the combination
\begin{equation}
I(f,\,\bm{\hat n}) = \, \langle h^{}_+(f,\,\bm{\hat n})\,h^\star_{+}(f,\,\bm{\hat n})\rangle  + \langle h^{}_\times(f,\,\bm{\hat n})\,h^\star_{\times}(f,\,\bm{\hat n})\rangle \,;
\end{equation}
note that this relates to the normalised logarithmic energy density $\Omega_{\rm GW}(f)$ as
\begin{equation}
\Omega_{\rm GW}(f) = \frac{4\pi^2f^3}{\rho_c}I(f)\,.
\label{omegatoI}
\end{equation}
We will leave the reconstruction of the polarisation components for future work.

\subsection{Detector Response}\label{ssec:DecRe}
The time stream $s_A(t,\bm{x_A})$ generated by a single, two-armed interferometer detector labelled $A$ at time $t$ and position $\bm{x_A}$ can be Fourier expanded to yield the signal as a function of frequency \citep[see e.g.][]{Cornish2001}:
\begin{equation}
\tilde{s}_{A}(f)=\int_S\, d\bm{\hat n}\,\sum_{P=+,\,\times}\,h^{}_P\,(f,\,\bm{\hat n})\,F_A^P\,(f,\,\bm{\hat n})\,e^{i2\pi f\,\bm{n\cdot x}_A}\, ,
\label{dectresp}
\end{equation}
where $F_A^P$ is the polarisation response function defined through the contraction of the polarisation tensor and the detector tensor as
\begin{equation}
F_A^P(\bm{\hat n}) = \frac{1}{2}\left(\bm{u}_A\otimes\bm{u}_A-\bm{v}_A\otimes\bm{v}_A\right)^{ij} e^P_{\,\,ij}(\bm{\hat n})\,,
\label{dectensor}
\end{equation}
where $\bm{u}_A$ and $\bm{v}_A$ are vectors describing the orientation of the two arms of the detector.

A signal that is composed of an incoherent superposition of many components, as in the case that we are focused on here, will vanish when averaged in time. To observe an incoherent GWB we therefore need to consider the integration of the square of the signal. This could be done by squaring the signal of a single detector but the Signal-to-Noise Ratio (SNR) level can be improved greatly by considering the {\sl cross-correlation} of two or more detectors under the assumption that their individual noise contributions are not correlated. We thus consider the signal of baselines defined by the vector $\bm{b} = \bm{x}_A-\bm{x}_B$
\begin{equation}
\begin{split}
d_b(f) &\equiv\, \langle \tilde{s}^{}_{A}(f)\tilde{s}^\star_{B}(f')\rangle\\
&=\int_{S^2} \!\!d\bm{\hat n}\,d\bm{n'}\!\!\!\!\! \sum_{P,P'=+,\,\times}\!\!\!\langle h^{}_Ph'^\star_{P'}\rangle \,F_A^P\,F_B^{P'}\,e^{i2\pi \, (f\,\bm{\hat{n}\cdot x}_A-f'\,\bm{\hat{n}'\cdot x}_B)}\,.
\end{split}
\label{corr}
\end{equation}
Expanding in the polarisation bases and inserting the relation between the second order moments of the strain and the Stokes parameters defined in (\ref{stokey}) we obtain an observing equation for the frequency domain signal of a single baseline labelled with subscript $b$
\begin{equation}
d_{b}(f) =  \int_S d\bm{\hat n} \,\sum_W \,\gamma^W_{b}(f,\,\bm{\hat n})\,W(f,\,\bm{\hat n})\,e^{i2\pi \,f\,\bm{\hat{n}\cdot b}}\,,
\label{dbf}
\end{equation}
where $W \equiv I$, $Q$, $U$, and $V$ and $\gamma^W_{\bm{b}}$ are known as overlap functions and are constructed through different combinations of the beam pattern functions of the two baseline detectors,
\begin{align}
\gamma^I_{b} (\bm{\hat n}) &= \frac{5}{8\pi} \left(F^+_AF^{+\star}_B+F^\times_AF^{\times\star}_B\right)\,,\\
\gamma^Q_{b} (\bm{\hat n}) &=  \frac{5}{8\pi} \left( F^+_AF^{+\star}_B-F^\times_AF^{\times\star}_B\right)\,,\\
\gamma^U_{b} (\bm{\hat n}) &=  \frac{5}{8\pi} \left( F^+_AF^{\times\star}_B+F^\times_AF^{+\star}_B\right)\,,\\
\gamma^V_{b} (\bm{\hat n}) &=-i  \frac{5}{8\pi} \left(F^+_AF^{\times\star}_B-F^\times_AF^{+\star}_B\right)\,.
\label{gammas}
\end{align}
The overlap functions contain all the information regarding the relative position of the detectors and their orientation with respect to a fixed gravitational wave polarisation basis. The normalisation of the functions is imposed by the condition that angular integral of $\gamma^W_{b}$ is unity in the case of two co-located and co-aligned detectors \citep{Allen1999}. In the following we fix our coordinate system and polarisation basis to the galactic coordinate frame. For Earth based detectors being considered here this means that the $\gamma^W_b$ will not be stationary. To avoid cumbersome notation we will make this time--dependence implicit in our equations but it will be useful to recall that the baseline vector is not stationary in galactic coordinates in that $\bm b\rightarrow \bm{b}(t)$. As we are considering only the intensity ($W=I$) of the strain here we will also drop the Stokes parameter label.

\begin{figure}
\centering
\includegraphics[width=\columnwidth]{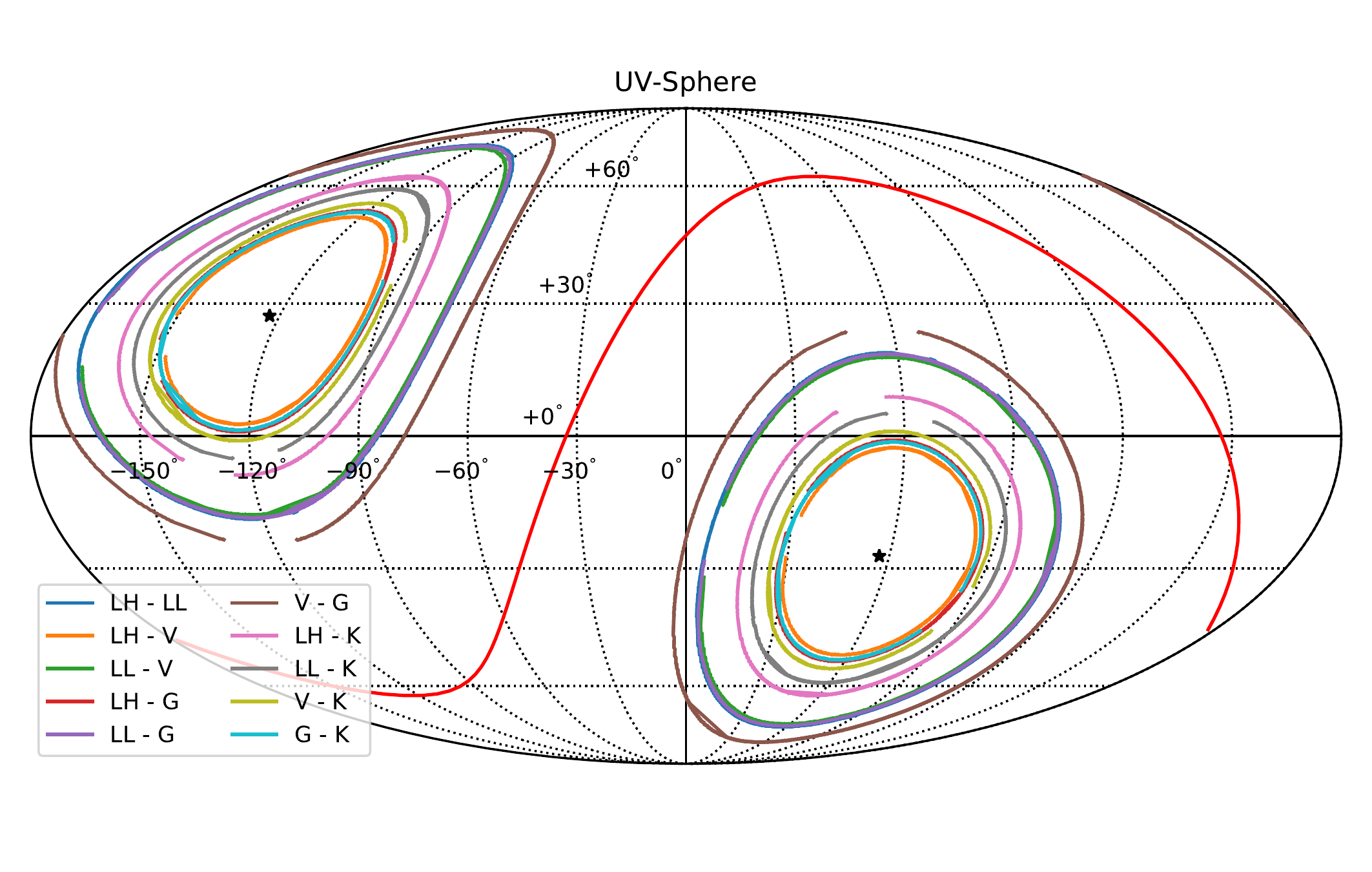}
\includegraphics[width=\columnwidth]{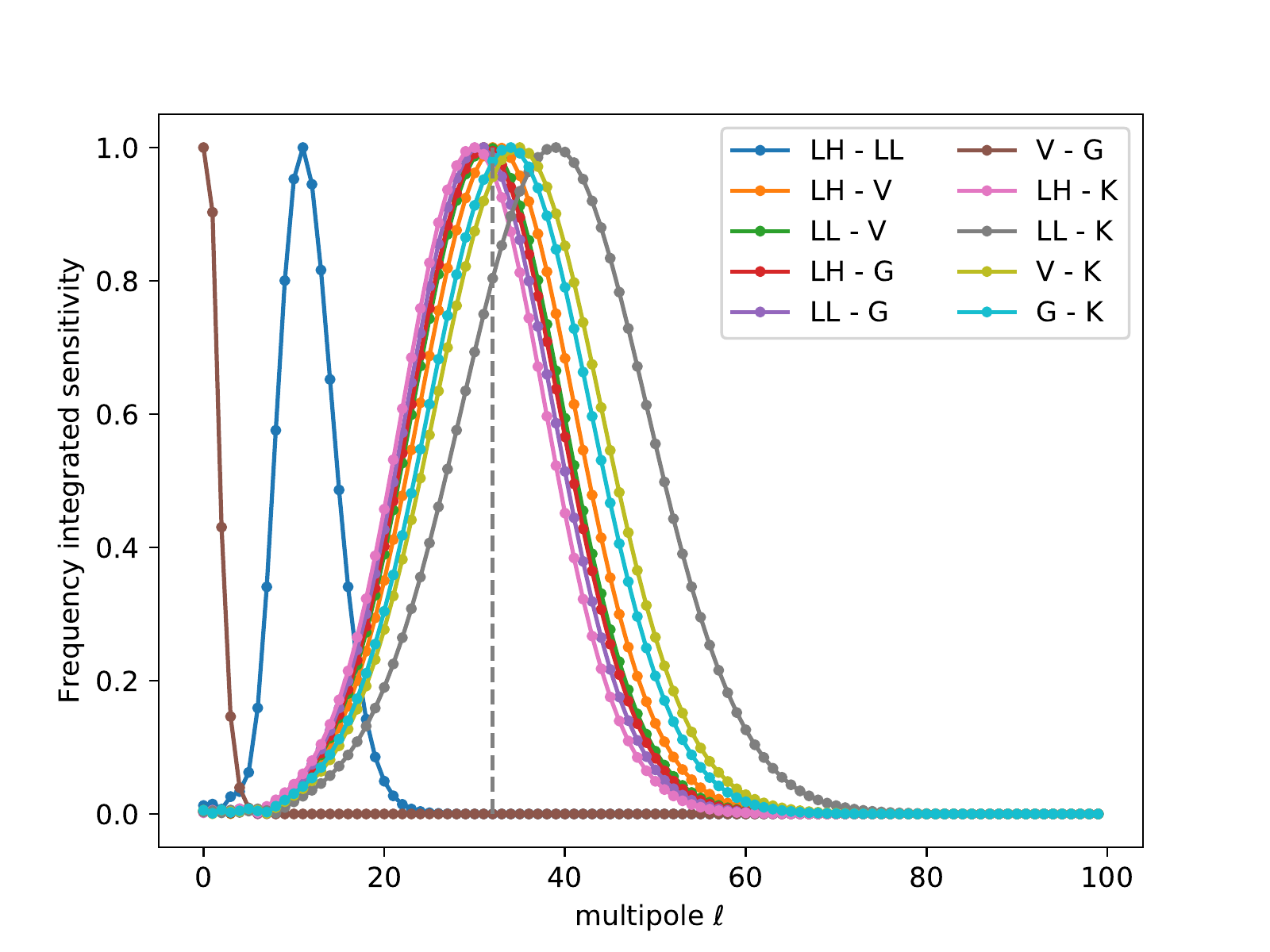}
\caption{Angular phase coverage ({\sl top}) of the 10 baseline case plotted in galactic coordinates. Each baseline track is plotted symmetrically about the celestial equator (red, solid curve). The tracks span 48 hours of simulated data following the segmentation and flagging of the actual ``O1'' LIGO data - only small gaps in the angular phase coverage of each baseline remain. An approximate view of the angular scale sensitivity  of each baseline is also shown ({\sl bottom}). The grey (dashed) vertical line at $\ell=32$ gives an indication of the limiting resolution in our reconstructed maps at our working pixelisation scale.}
\label{fig:uvw}
\end{figure}

It is useful at this stage to consider the observing characteristics that are particular to gravitational wave detectors. Although we will develop a map--maker that solves the sky directly in the coordinate domain it is convenient to expand the observations in spherical harmonics since the cross--correlation signal measures the Fourier domain directly. This is analogous to radio interferometry where the measurement can be most naturally represented in the ``$uv$--plane''. In the case of gravitational wave detectors however, the ``beam'' with which we convolve the sky is not compact and is also symmetric with respect to the pointing axis - the detectors are equally sensitive to gravitational waves of opposite momenta. This means that no flat--sky approximation can be used and a gravitational wave measurement can be represented on a ``$uvw$--sphere''.

To introduce the spherical harmonic basis we can expand the intensity field over basis functions $Y_{\ell m} (\bm\hat n)$ in terms of coefficients $a_{\ell m}$ \citep{Mitra2008,Gair:2014rwa}
\begin{equation}
I(f,\bm{\hat n}) =  E(f) \sum_{LM}a_{\ell m}Y^\star_{\ell m}(\bm{\hat n})\,,
\end{equation}
where we have also separated out the spectral dependence by introducing the function $E(f)$.

Inserting this into (\ref{dbf}) and making use of the spherical harmonic expansion  of the plane wave
\begin{equation}
	e^{i \bm{k}\cdot\bm{x}} = 4\pi\sum_{\ell m} i^\ell\,j_\ell(kx)\,Y^{}_{\ell m}(\bm{\hat k})Y^\star_{\ell m}(\bm{\hat x})\,,
\end{equation}
where $j_\ell(x)$ are spherical Bessel functions, we can obtain the observing equation for the directly observed ``$uvw$''--sphere harmonic coefficient $d^b_{\ell m}$ defined via 
\begin{equation}
d_b(f) = E(f)\sum_{\ell m}d^b_{\ell m}Y^\star_{\ell m}(\bm\hat b)
\end{equation}
as
\begin{equation}
d^b_{\ell m}=  i^\ell 4\pi \,E(f) \sum_{\ell'm'LM} j^{}_\ell(2\pi f b)a^{}_{\ell 'm'}  {\cal K}^{LM}_{\ell m,\ell'm'}\,.
\label{dlmK}
\end{equation}
The coupling kernel ${\cal K}^{LM}_{\ell m,\ell'm'}$ incorporates the Gauntt integral
\begin{align}
{\cal K}^{LM}_{\ell m,\ell'm'} &= \gamma^{}_{LM}\int_{S^2}d\bm{\hat n}\, Y_{LM}(\bm{\hat n})\,Y_{\ell m}(\bm{\hat n})\,Y_{\ell' m'}(\bm{\hat n})\,,\\
\begin{split}
&= \gamma^{}_{LM}\sqrt{\frac{(2L+1)(2\ell+1)(2\ell'+1)}{4\pi}} \begin{pmatrix} L & \ell & \ell' \\ 0 & 0 & 0\end{pmatrix} \, \times\\
&\begin{pmatrix} L & \ell & \ell' \\ M & m & m'\end{pmatrix}\,,
\end{split}
\label{eq:klm}
\end{align}
where the terms in brackets indicate $3j$ coefficients and the overlap function $\gamma(\bm\hat n)$ has also been expanded in spherical harmonics.

The projection of three--dimensional plane waves onto the two--dimensional sphere, together with a non--uniform, non--compact weighting of the sky, introduces the coupling between $\ell$, $\ell'$, and $L$ in the spherical harmonic expansion. This expansion is useful when considering how spherical modes are reconstructed by the observations. One aspect that is peculiar to gravitational wave detectors is that they are equally sensitive to modes travelling in opposite directions - there is no possibility of beam forming since gravitational waves cannot be focused using detectors on feasible scales. This implies that the overlap functions are symmetric and therefore the detectors are insensitive to a mode that is odd with respect to the same axis. This implies that $\gamma_{\ell m}=0$ for odd $\ell$ and one might expect that this in turn means the observations cannot reconstruct $a_{\ell m}$ modes with odd $\ell$ since a general rotation is only couples $m$ and is orthogonal in $\ell$. This is not the case however since the projection of the plane waves onto the sphere breaks the orthogonality - different wave-vectors $\bm k$ contribute to each projected $\ell$ mode and this breaks the symmetry implied by the overlap functions. This is easily seen in (\ref{dlmK}) where we can consider the observation of an odd mode on the sky $\ell' = 2p+1$ through an even mode of the overlap function $L=2q$ with $p,\,q \in \cal {\mathbb{Z}}^+$. In that case the $3j$ coefficients are non-zero in the range $|2(p-q)-1| \leq \ell \leq 2(p+q)+1$ implying that a measurement at ``$uvw$'' point $d^b_{\ell m}$ does not vanish for odd $\ell'$.

Given sufficient (angular) phase coverage of the ``$uvw$''--sphere it should be possible to reconstruct all spherical modes on the sky within a range of angular scales - this should be a well--conditioned inversion problem. For an Earth based detector the phase coverage is provided by the Earth's rotation and the latitudinal positioning of the detectors involved in the baselines. Table~\ref{tab:dects} lists the position and orientation of five detectors considered in this work. The LIGO and VIRGO detectors are operational and currently observing but only data from the first LIGO observing run (O1) has been released\footnote{\url{http://losc.ligo.org}}. This includes observation from only the two LIGO detectors (LH and LL). The remaining two detectors are not operational yet but we have included them in some of our simulated combinations in order to test our algorithm on an extended set of baselines.

\begin{table}
\begin{center}
\caption{Latitude, longitude and the orientation angle $\alpha$ in degrees of the five, currently operating and future, detectors considered as part of various combinations in this work. Specifically, $\alpha$ is the angle between the local parallel and the bisector of the aperture of the interferometer \citep[see also][]{Seto2008}.}
\label{tab:dects}
\begin{tabular}{ l |l  l l l }
\hline
Detector & Label  & Latitude     &  Longitude     &   $\alpha$  \\
\hline
\hline
 LIGO Hanford & LH & 46.4 & -119.4 & 171.8\\ 
 LIGO Livingston & LL & 30.7 & -90.8 & 243.0\\  
 VIRGO & V & 43.6 & 10.5   & 116.5\\
 Kagra & K & 36.3 & 137.2 & 225.0 \\
 GEO600 &G & 48.0 & 9.8 & 68.8 \\
\hline
\end{tabular}
\end{center}
\end{table}

Figure~\ref{fig:overlap} shows the overlap functions, in galactic coordinates and at arbitrary time, for the single LIGO (LL-LH) baseline and the sum of the overlap functions of the ten baseline set given by the full five detector combination being considered. The addition of baselines means the sky signal is integrated more extensively at each observation but the weighting of the signal remains inhomogeneous. The restricted range of latitudes of the detectors also affects the coverage on the sky. The combination of extensive but inhomogeneous weighting and limited range in latitudes means that, on time scales shorter than 24 hours, the problem of sky reconstruction is ill--conditioned. However, as we will show, integrating for sufficiently long time scales improves the conditioning considerably.

Figure~\ref{fig:uvw} shows a more useful representation of the coverage of the ten baseline set. The track of each baseline vector direction can be plotted on the sky together with an integrated measure of the sensitivity to different angular scales on the sky. To obtain the sensitivity curves we integrate the spherical Bessel function for each baseline in frequency with a Gaussian weight centered at 200 GHz with deviation $\sigma=50$ GHz. This roughly emulates the sensitivity of current LIGO-type detectors and allows us to estimate the range of scales each individual baseline is sensitive to. The coupling of modes significantly complicates the interpretation of the baseline resolution in terms of an overall resolution scale of any final map. For example, the 3$j$ coefficients in (\ref{eq:klm}) correlate a multipole $\ell$ with multipoles $\ell'$ in the range $|\ell-\ell'| \le 2\ell$.

As detailed in Section~\ref{sec:res} we will use the ``O1'' release LIGO time stream data \cite{TheLIGOScientific:2016agk} as a template for the segmentation and flagging of all our simulated runs. The data contains gaps for periods where either detectors forming each baseline were off--line or for periods when the data is flagged in such a way that it cannot be used for analysis. Some of these gaps can be seen in the baseline tracks in Figure~\ref{fig:uvw} that cover a single 48 hour period. Since the gaps are not expected to be particularly correlated with diurnal phase, their overall effect should diminish as integration time increases beyond a few days.

\section{Maximum--likelihood sky reconstruction}\label{sec:MaMa}

\begin{figure}
\centering
\includegraphics[width=\columnwidth]{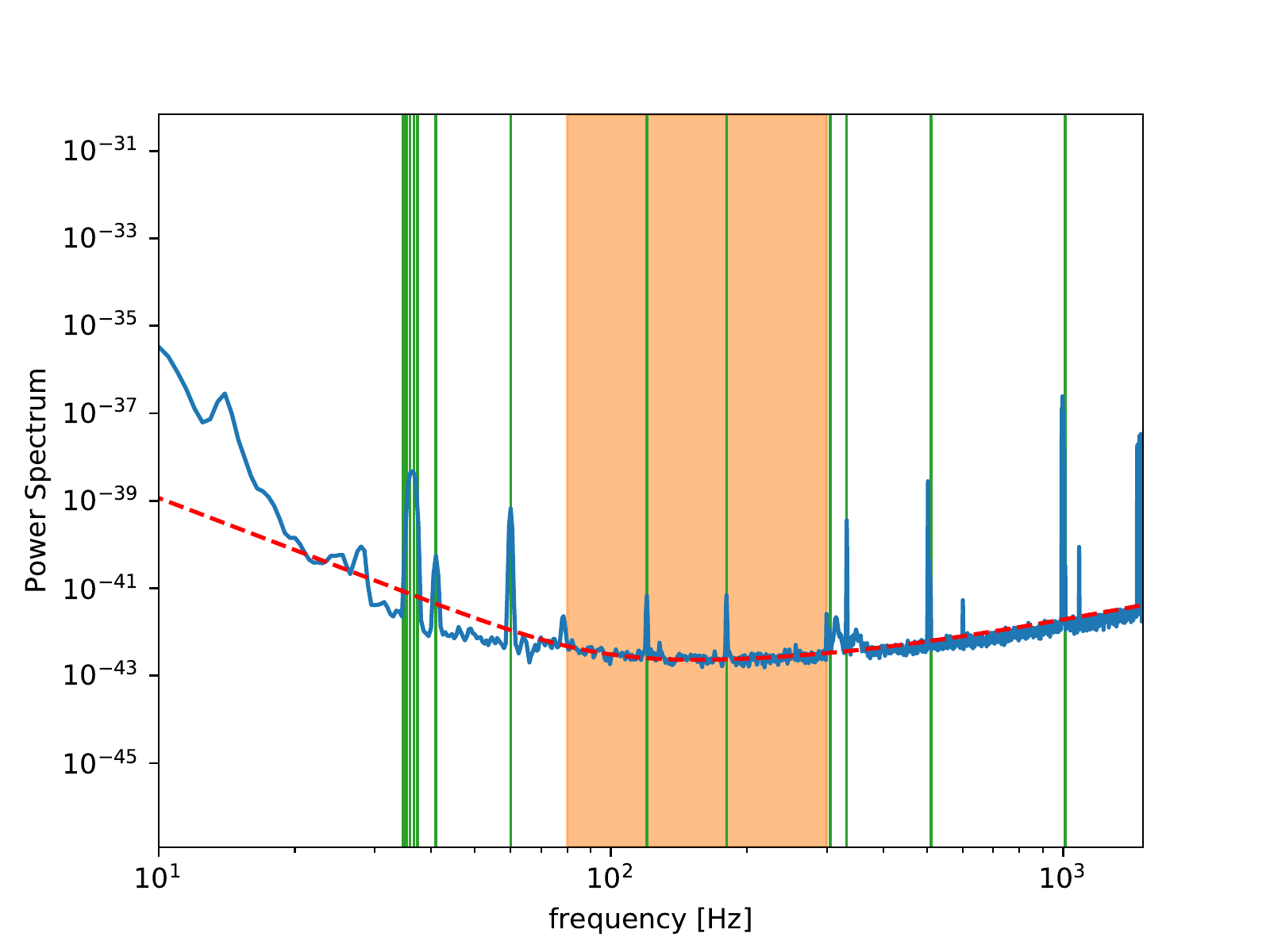}
\caption{LIGO LL, smoothed power spectrum for a single 60 second time segment. The vertical lines show the positions of the notches applied to the frequency data in order to remove known harmonics. The dashed curve shows the model fit to the spectrum within the frequency range shown in the shaded area. The fit is carried out for every 60 second time segment to all detector time streams. When fitting the spectrum the data is gapped around the notched frequencies to avoid a biasing the weighting function used in the map--making and in the estimate of the pixel noise covariance matrix. The frequency range adopted here is a conservative one that limits $1/f$ and $f$ tail contributions that can fluctuate significantly between segments of the data. }
\label{ligo_psd}
\end{figure}

\begin{figure*}
\centering
\begin{minipage}{.32\textwidth}
  \centering
  \includegraphics[width=0.95\linewidth]{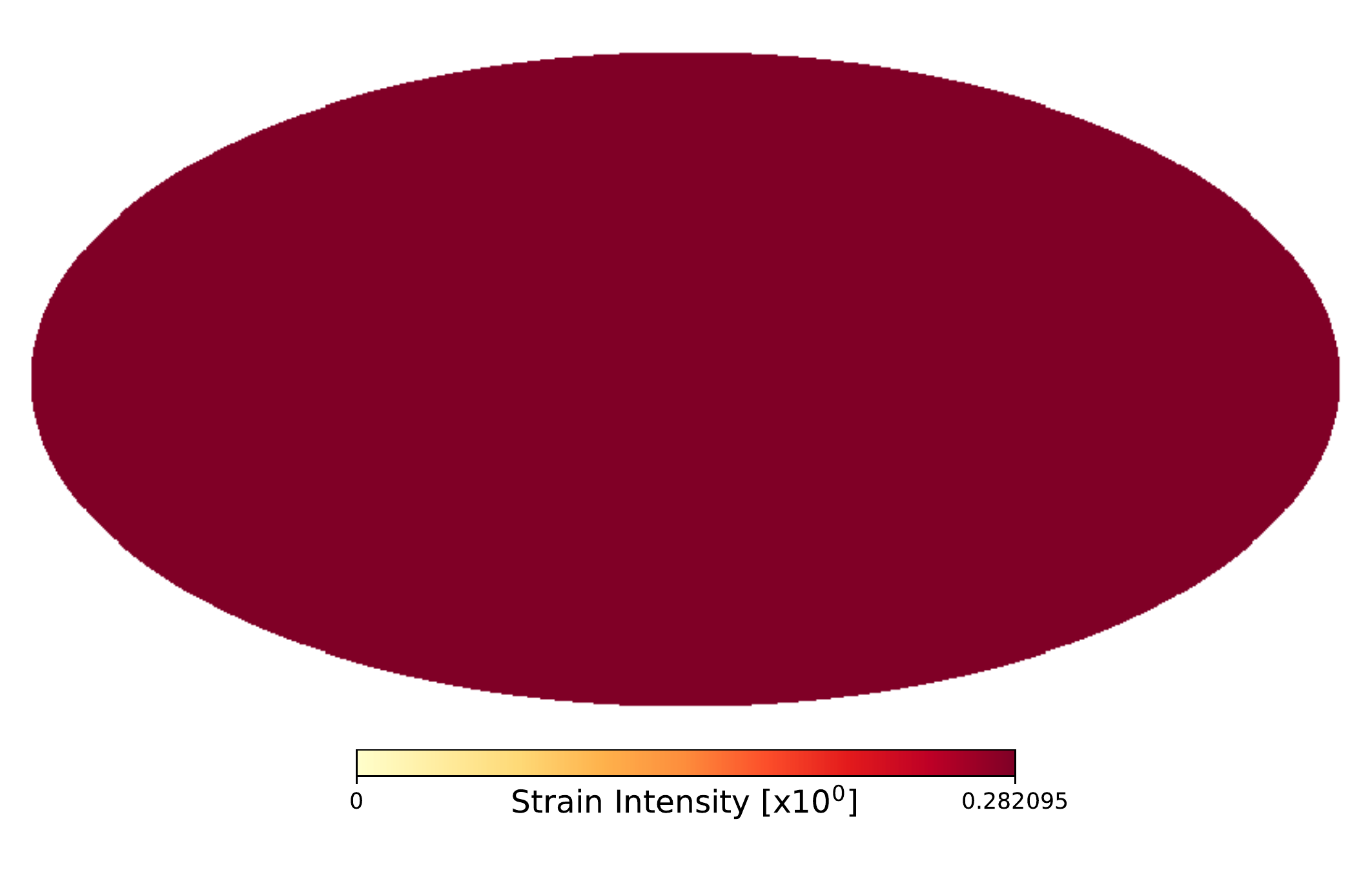}
\end{minipage}%
\begin{minipage}{.32\textwidth}
  \centering
  \includegraphics[width=0.95\linewidth]{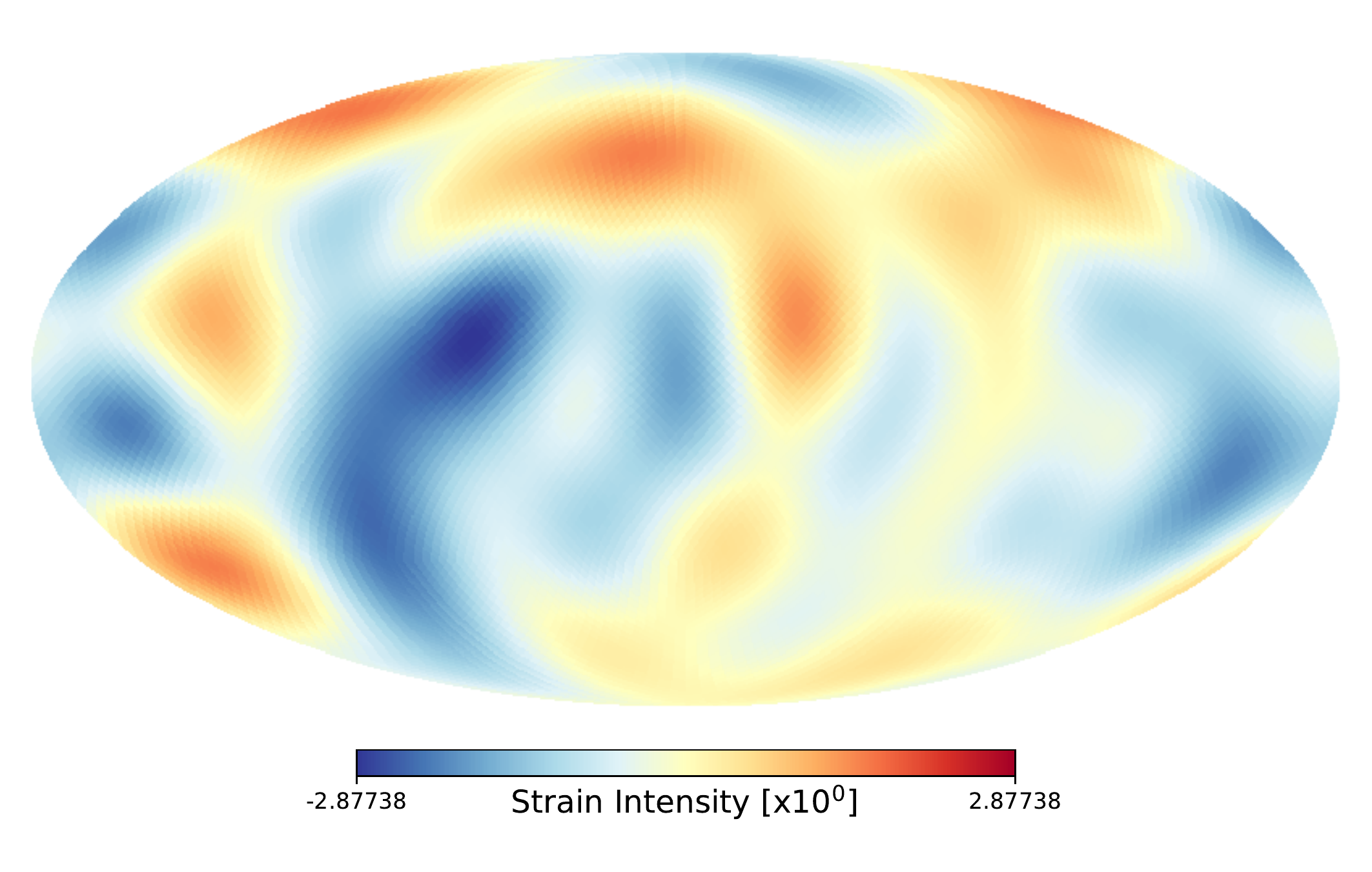}
\end{minipage}
\begin{minipage}{.32\textwidth}
  \centering
  \includegraphics[width=0.95\linewidth]{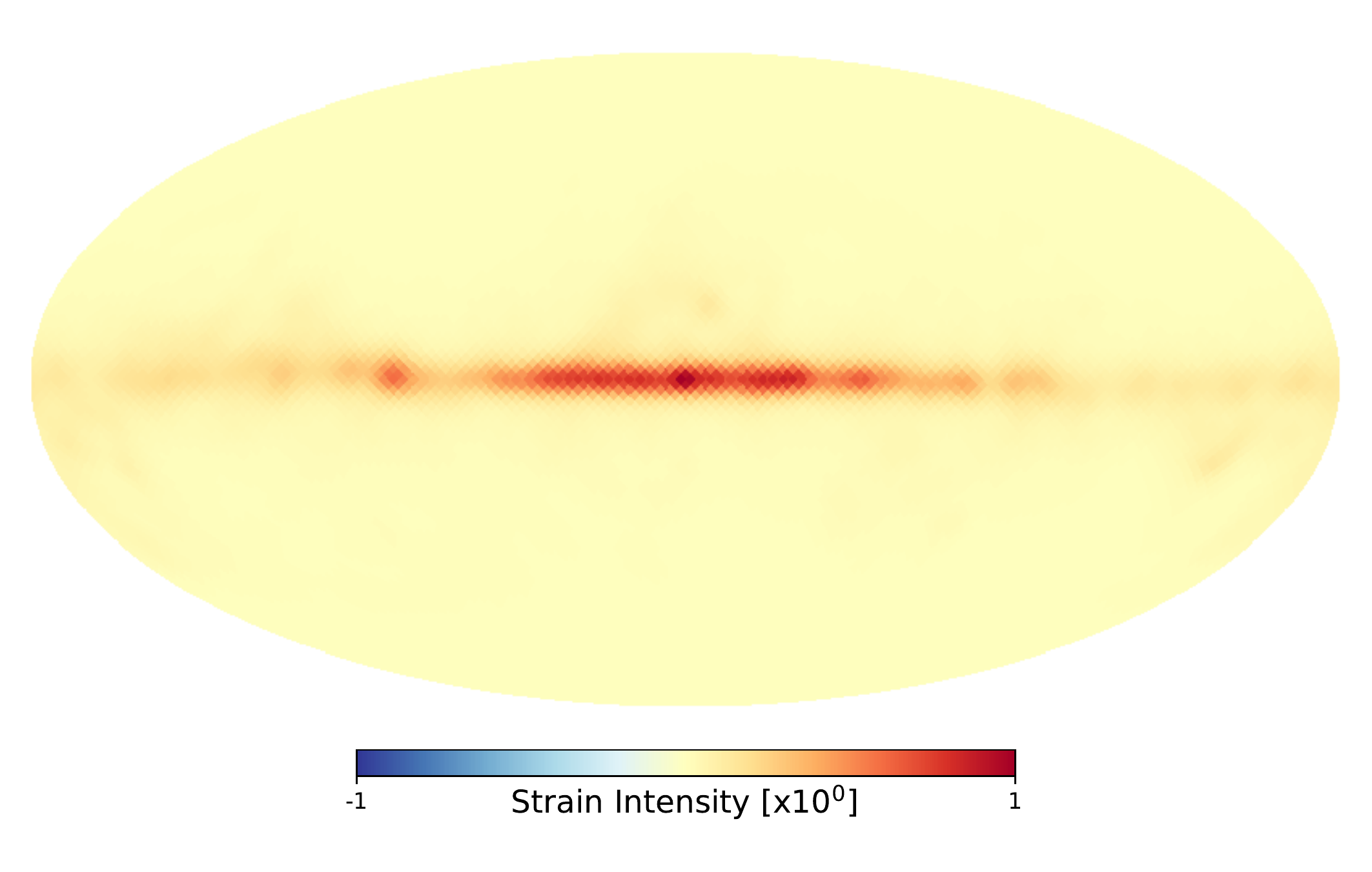}
\end{minipage}
\begin{minipage}{.32\textwidth}
  \centering
  \includegraphics[width=0.95\linewidth]{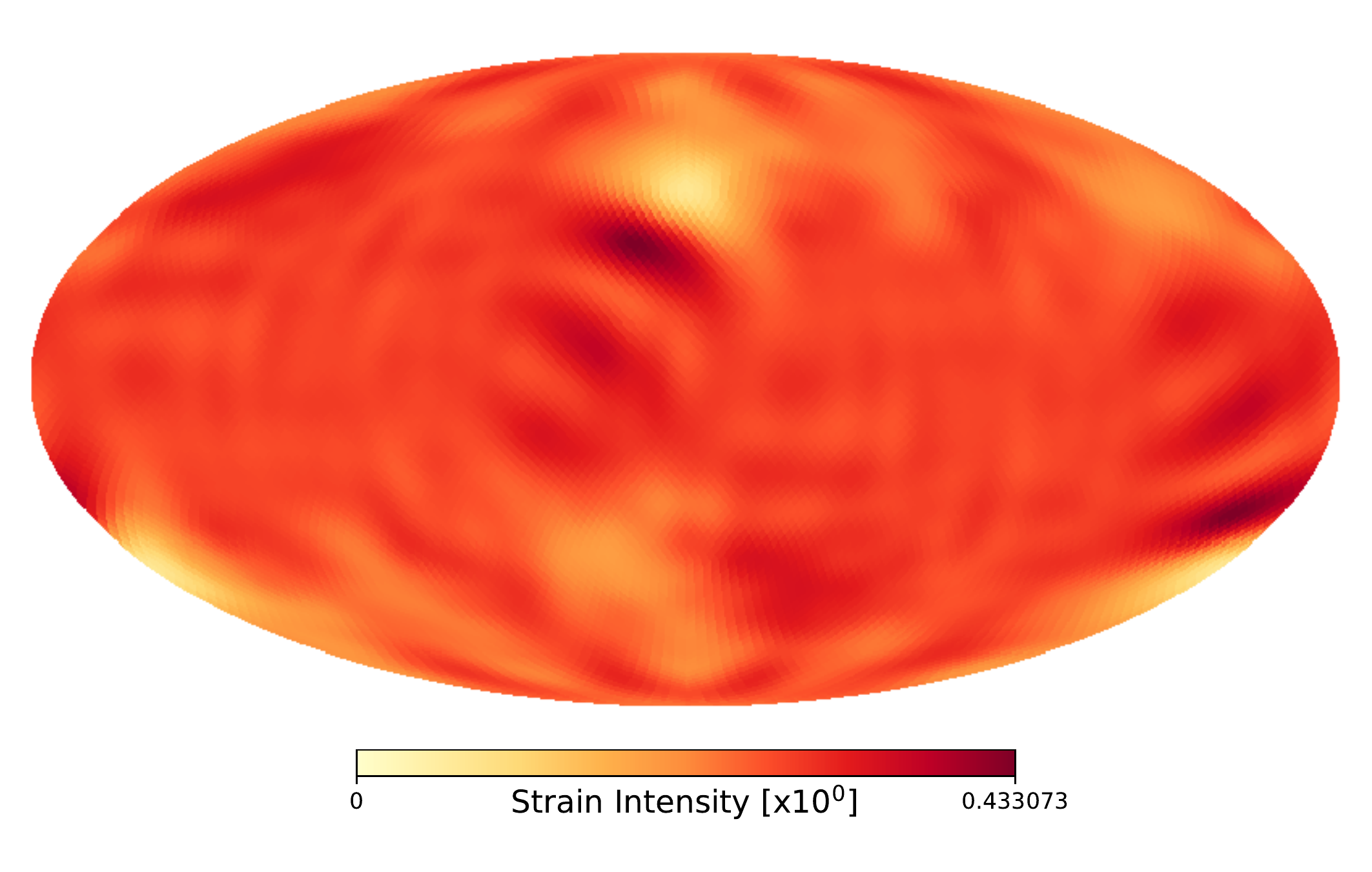}
\end{minipage}%
\begin{minipage}{.32\textwidth}
  \centering
  \includegraphics[width=0.95\linewidth]{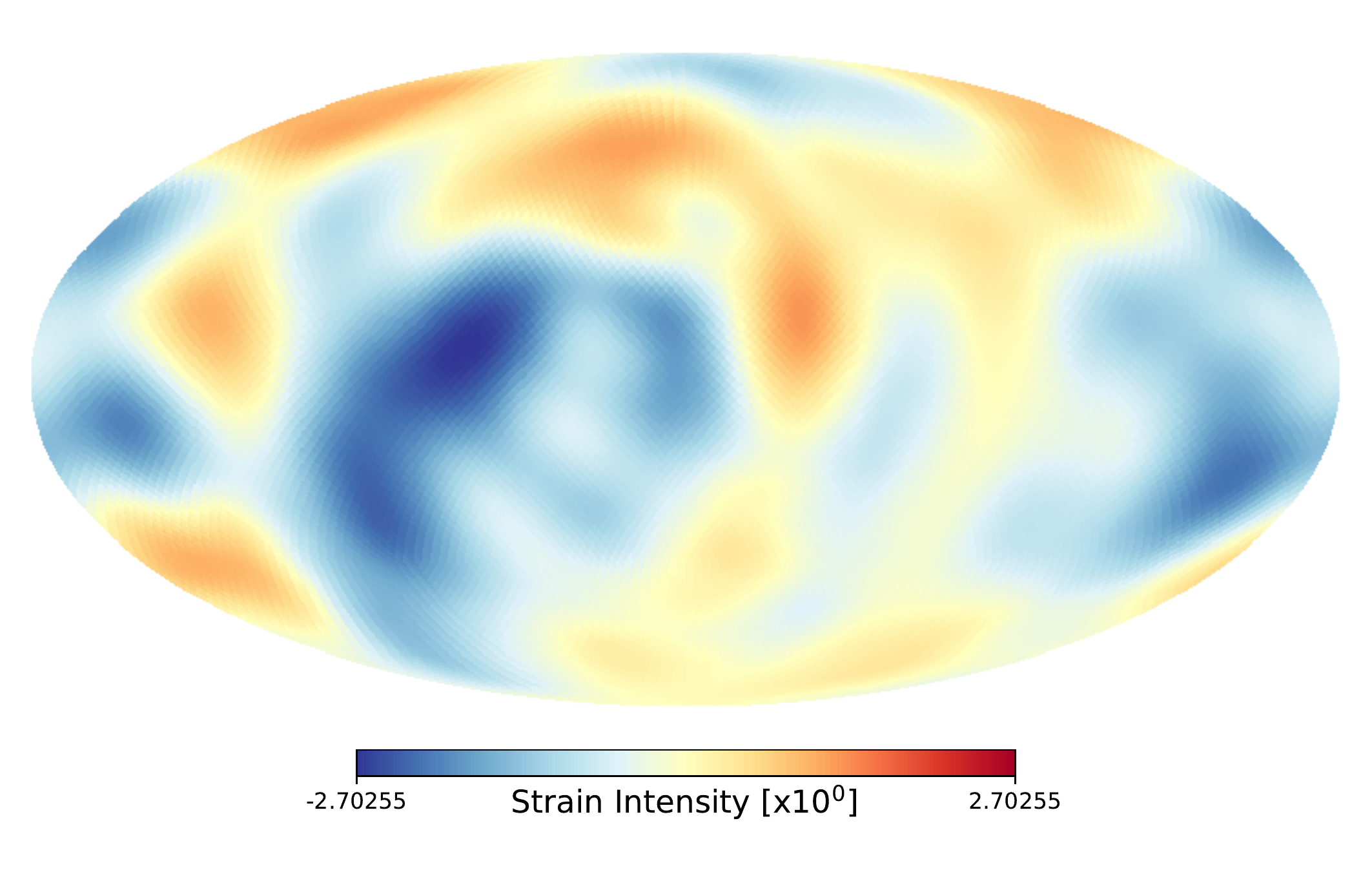}
\end{minipage}
\begin{minipage}{.32\textwidth}
  \centering
  \includegraphics[width=0.95\linewidth]{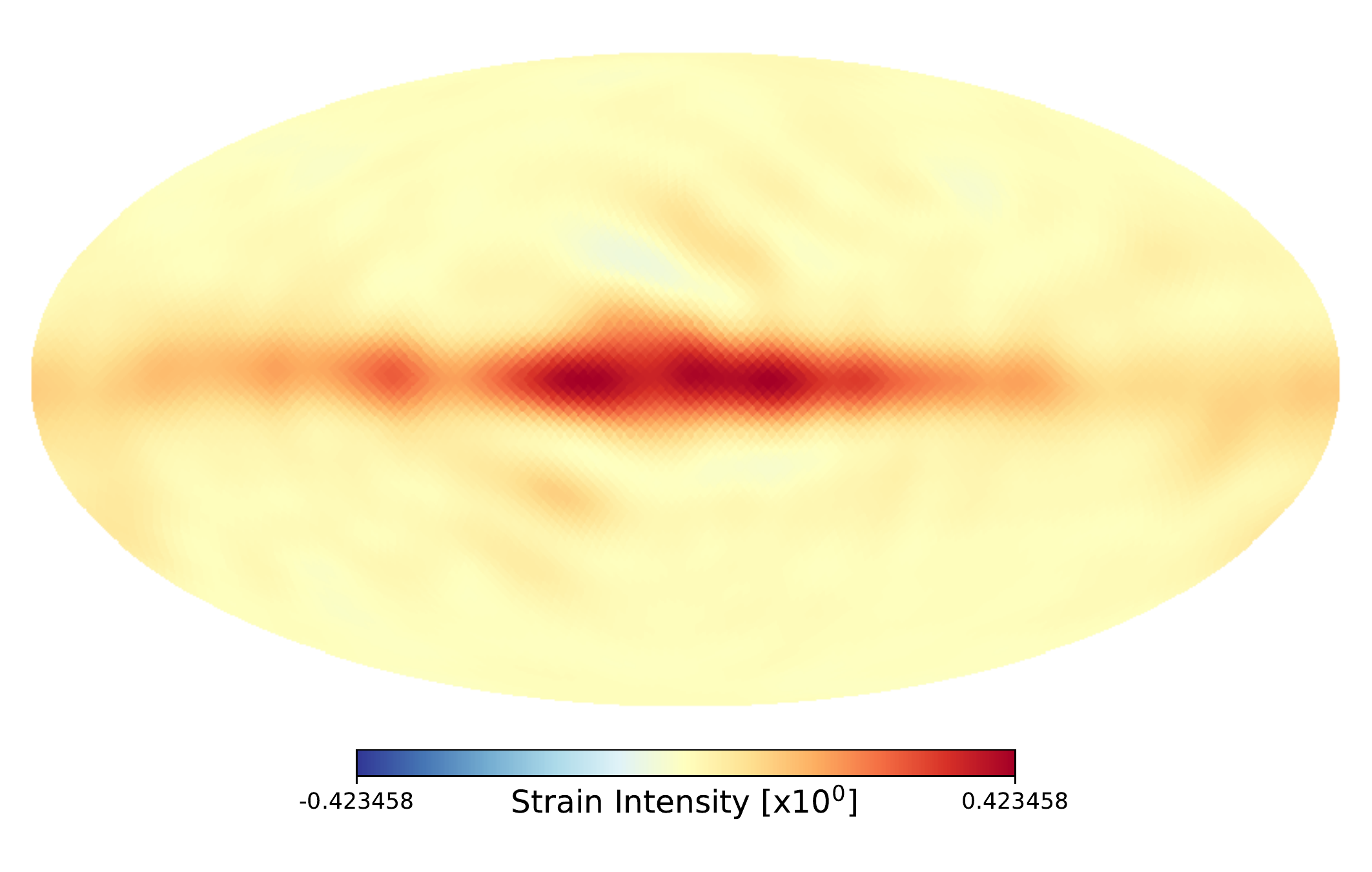}
\end{minipage}
\caption{Input ({\sl top}) and output ({\sl bottom}) maps for the high SNR cases. The results are for the analysis of 24 hours, simulated data, mimicking the LIGO ``O1'' release data structure for the three baseline (LIGO-VIRGO) case. Left: The input monopole map with amplitude $1/2\sqrt{\pi}\sim 0.282$ corresponding to a monopole $a_{00}=1$. The output map recovers an average $\mu = 0.278$. The non-trivial ``$uvw$'' coverage results in significant correlations in the final map. Middle: The ``Gaussian'' statistically isotropic map case. All $\ell \le 8$ modes are recovered accurately. Right: The anisotropic ``Galaxy'' case. The resolution limit of the observations is most obvious in this case.}
\label{fig:hsnr}
\end{figure*}

In this section we briefly review the generalised problem of finding the maximum-likelihood solution for a map from the time domain data of a scanning experiment \citep[see e.g.][]{Borrill:1999kt}. This has been considered by a number of authors \citep{Thrane2009,Romano2015,Abbott2017d} for gravitational wave telescopes and these have made different choices with respect to solving for a map in sky coordinates (pixels) or in the Fourier domain (spherical harmonic coefficients). Usually, in interferometry, the choice has been to obtain a Fourier domain solution since the measurements directly constrain this domain \cite[see e.g][for an application to CMB interferometry]{Myers:2002tn}. The reconstruction of a sky ``image'' is then left as a separate inversion problem. The non--trivial coupling of spherical modes implied by (\ref{dlmK}) indicates that adopting this approach for gravitational wave detectors may have significant drawbacks. In particular, we don not have in this case, an isotropic kernel for a primary beam describing the coupling of individual $uv$ points. Therefore, limiting the number of spherical modes in the solution in a piecewise fashion could make the problem more ill-conditioned than it actually is. This is the main motivation for the alternative approach we adopt here where we solve for the map directly in the pixel domain.

The map-making problem can be stated as an inversion of a general observing equation
\begin{equation}
\bm{d} = \bm{A}\,\bm{s} + \bm{n}\,,
\label{eq:observe}
\end{equation}
where $\bm{d}$ is a vector containing the observations, the linear operator $\bm{A}$ projects the signal ``map'' $\bm{s}$ into the domain of the observations, and $\bm{n}$ is a noise component. The observations usually span the time domain and constitute a data time stream whilst the signal component spans a separate ``pixel'' domain. The noise component spans the same domain as the observations. Under the assumption that the noise component is a zero--mean, Gaussian variate with covariance $\bm{N} = \langle \bm{n}\otimes\bm{n}\rangle$ the maximum--likelihood solution for the signal map is obtained by minimizing the log likelihood
\begin{equation}
\ln L = \frac{1}{2}\left[ (\bm{d}-\langle \bm{d}\rangle)^\dagger \bm{N}^{-1} (\bm{d}-\langle \bm{d}\rangle) + \Tr \ln \bm{N}\right]\,,
\end{equation}
with respect to the signal map $\bm s$, giving the closed--form solution for the estimated map
\begin{equation}
	\bm{\tilde s} = \left(\bm{A}^\dagger \bm{N}^{-1} \bm{A}\right)^{-1} \bm{A}^\dagger \bm{N}^{-1} \bm{d}\,.
    \label{eq:soln}
\end{equation}

For gravitational wave data it is convenient to Fourier transform the observed time stream so that $\bm{d}$ spans the frequency domain. With our choice of $\bm{s}$ spanning the sky pixel domain (\ref{eq:observe}) then becomes
\begin{equation}
d_f = \sum_p A_{fp}s_p + n_f\,,
\end{equation}
where $p$ runs over all pixels on the sky. In the frequency domain the noise contribution takes on a particularly simple form with 
\begin{equation}
\langle n_f^{}n_{f'}^\star\rangle \equiv N_{ff'} =  \delta(f-f') P(f)\,. 
\end{equation}

We use the Healpix\footnote{\url{http://healpix.sourceforge.net}} \citep{Gorski:2004by} hierarchical pixelisation scheme which discretises the sky into equal area elements. The equal area scheme allows us to discretise the observation equation (\ref{dbf}) for a single baseline vector $\bm{b}$ as
\begin{equation}
d^\tau_f = E_f \Delta_f \frac{4\pi}{N_{\rm pix}}\sum_p\, I_p \,\gamma^\tau_p\, e^{i2\pi f\, \bm{b^\tau}\cdot \bm{\hat p}} + n_f\,,
\label{eq:dtf}
\end{equation}
where $N_{\rm pix}$ is the total number of sky pixels, $\Delta_f$ is the discretisation interval in frequency, $I_p$ is the intensity on the sky at pixel $p$ with unit direction $\bm{\hat{p}}$, and $\gamma_p^\tau$ is the discretised overlap function. We have added a superscript $\tau$ to indicate that this is the observation at a particular time frame $\tau$ defined by a fixed pointing with respect to the sky frame. This highlights that in the galactic frame adopted here the baseline vector and overlap function are both rotating with respect to the sky.

We can now identify the operation $\bm{A}^\dagger \bm{N}^{-1} \bm{d}$ required in (\ref{eq:soln}) with the discretised form
\begin{equation}
 \bm{A}^\dagger \bm{N}^{-1} \bm{d} \to z_p=\Delta_f \frac{4\pi}{N_{\rm pix}}\sum_f \,\gamma^\tau_p\,\frac{E^{}_fd^\tau_f}{P_f}\, e^{i2\pi f\, \bm{b}^\tau\cdot \bm{\hat p}}\,.
\end{equation}
It is important to note that the reality of the observed time streams implies the condition $d^\star_f = d^{}_{-f}$ such that the operation can be carried out as a some over the positive frequency domain only
\begin{equation}
\begin{split}
	z_p=\Delta_f \frac{8\pi}{N_{\rm pix}}\sum_{f=0}^\infty \,\gamma^\tau_p\,\frac{E^{}_fd^\tau_f}{P_f}\, \left[ \cos(2\pi f\, \bm{b}^\tau\cdot \bm{\hat p})\mathbb{R}(d^\tau_f)\right.-\\
    \left.\sin(2\pi f\, \bm{b}^\tau\cdot \bm{\hat p})\mathbb{I}(d^\tau_f) \right]\,.
    \end{split}
\end{equation}
Similarly we can identify the operation $\bm{A}^\dagger \bm{N}^{-1} \bm{A}$ in (\ref{eq:soln}) as
\begin{equation}
	\begin{split}
	M_{pp'} = \Delta_f^2 \frac{16\pi^2}{N_{\rm pix}}\sum_{f=0}^\infty \,\gamma^\tau_{p^{}}\,\gamma^\tau_{p'}\,\frac{E^2_f}{P_f}\,\cos\left[2\pi f\, \bm{b}^\tau\cdot (\bm{\hat p}-\bm{\hat{p}})\right]\,,
    \end{split}
\end{equation}
giving the maximum-likelihood map estimate
\begin{equation}
	{\tilde s}_p = \sum_{p'} \,M^{-1}_{pp'} z^{}_{p'}\,.
    \label{eq:ml_map}
\end{equation}
When more than a single pointing time frame $\tau$ is present and when the observations cover multiple baselines $b$ the expressions are modified by summing over the individual contributions in such a away that 
\begin{equation}
z_p = \sum_{\tau,b} z^{\tau,b}_p\,,
\end{equation}
and
\begin{equation}
M_{pp'} = \sum_{\tau,b} M_{pp'}^{\tau,b}
\end{equation}
where the superscripts denote the individual contributions to each time frame and baseline. Each of these will involve a rotation of the baseline dependent overlap function $\gamma^{\tau, b}_p$ maps and baseline vectors $\bm{b}^\tau$ to the sky frame. We use the {\tt QPoint}\footnote{\url{http://github.com/arahlin/qpoint}} library of \cite{Rahlin} to apply the location dependent transformation from detector to sky frames. The power spectrum of the noise will also, in general, be different for separate time frames $\tau$ and this is also included implicitly in the operation by re-estimating the noise for each time frame as discussed below.  

\begin{figure*}
\centering
\begin{minipage}{.32\textwidth}
  \centering
  \includegraphics[width=0.95\linewidth]{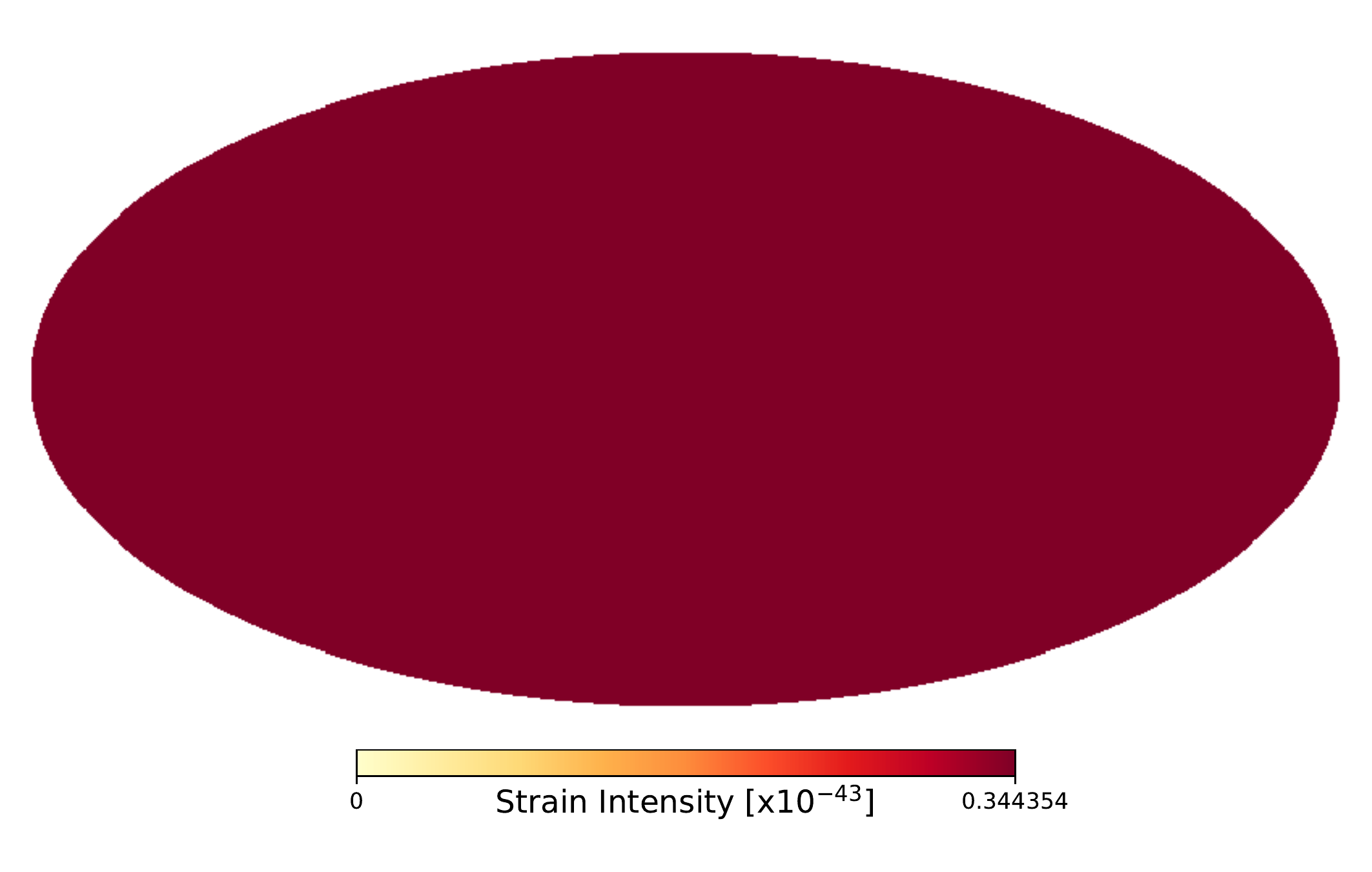}
\end{minipage}%
\begin{minipage}{.32\textwidth}
  \centering
  \includegraphics[width=0.95\linewidth]{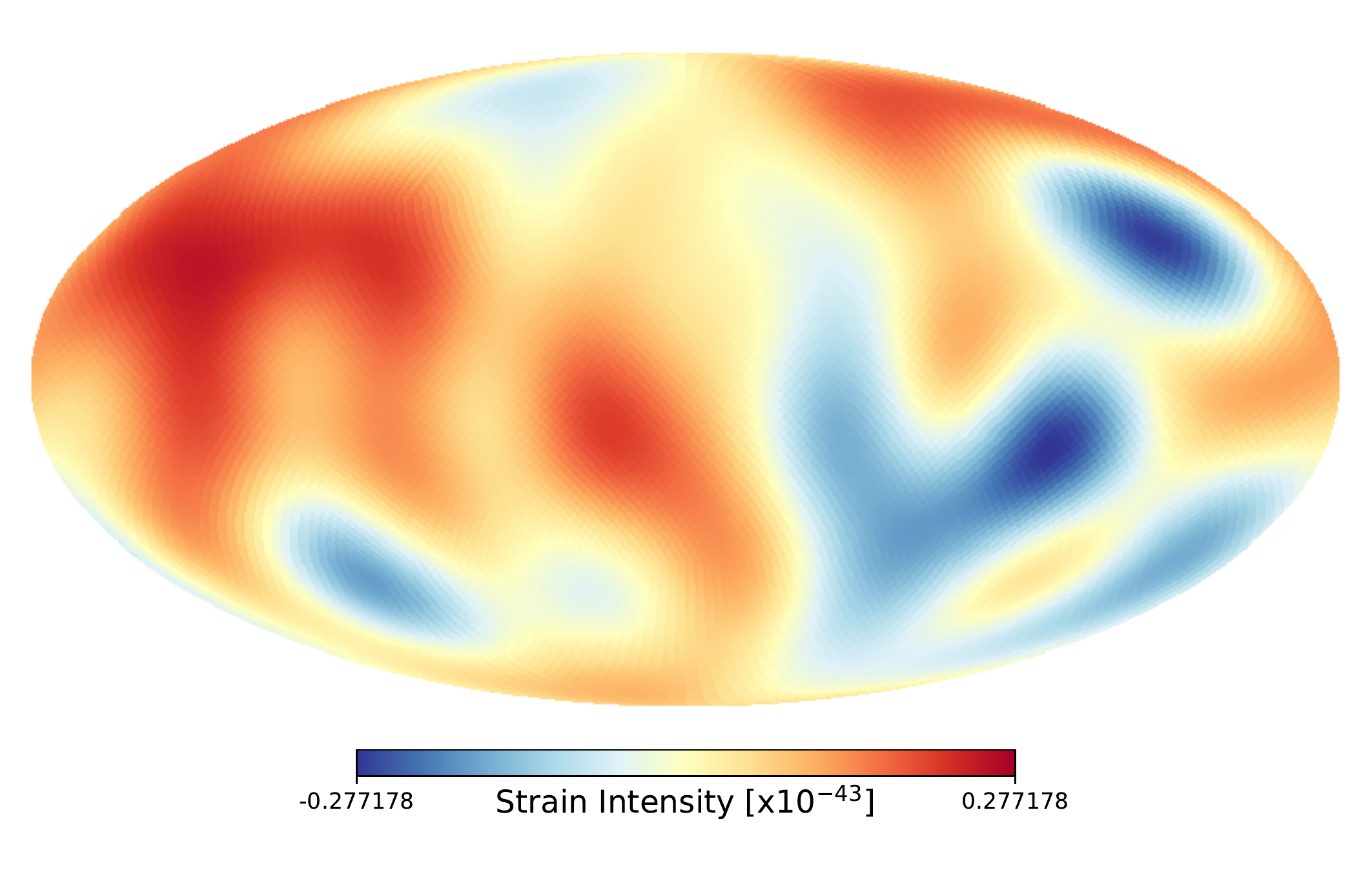}
\end{minipage}
\begin{minipage}{.32\textwidth}
  \centering
  \includegraphics[width=0.95\linewidth]{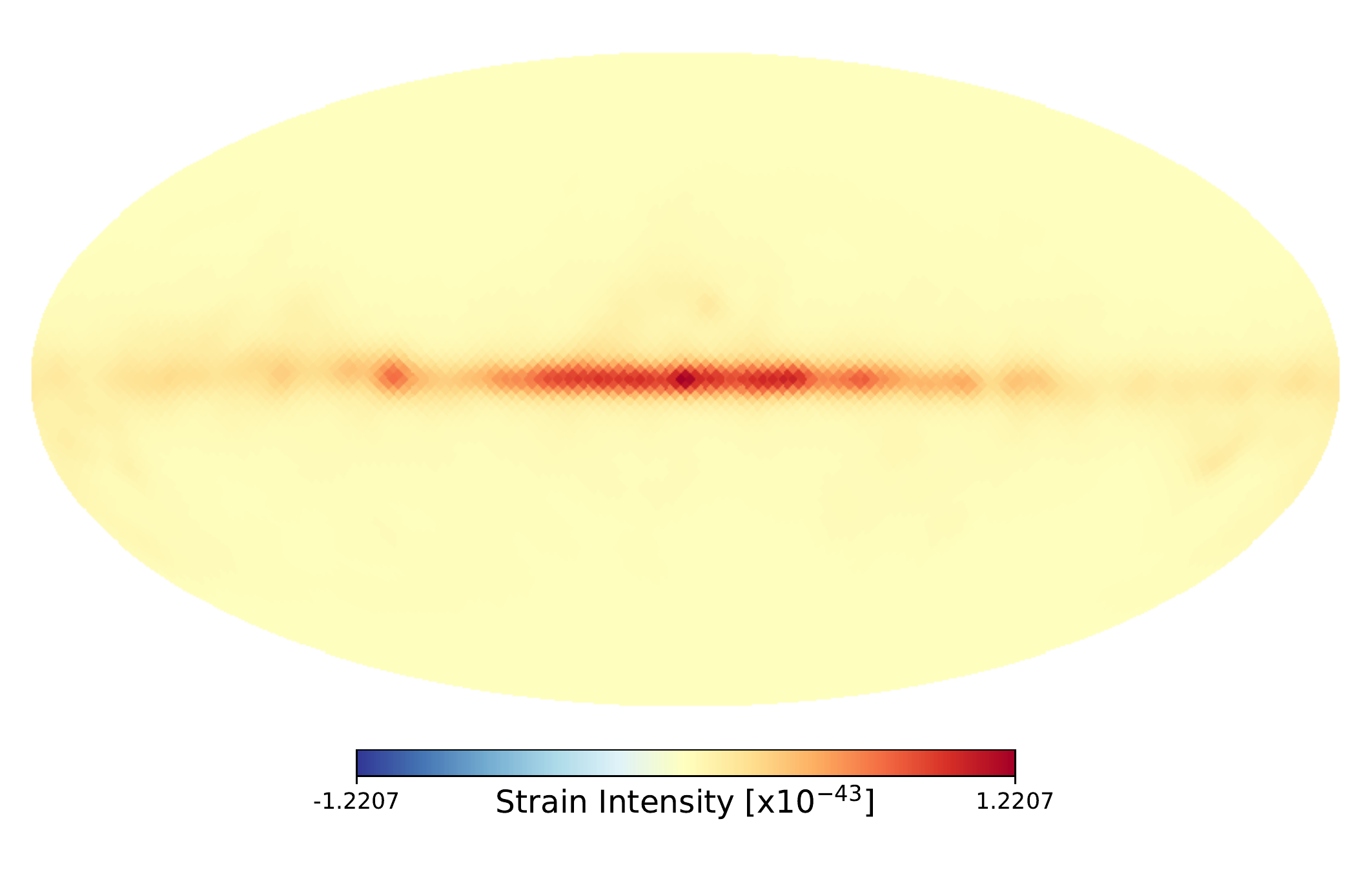}
\end{minipage}
\begin{minipage}{.32\textwidth}
  \centering
  \includegraphics[width=0.95\linewidth]{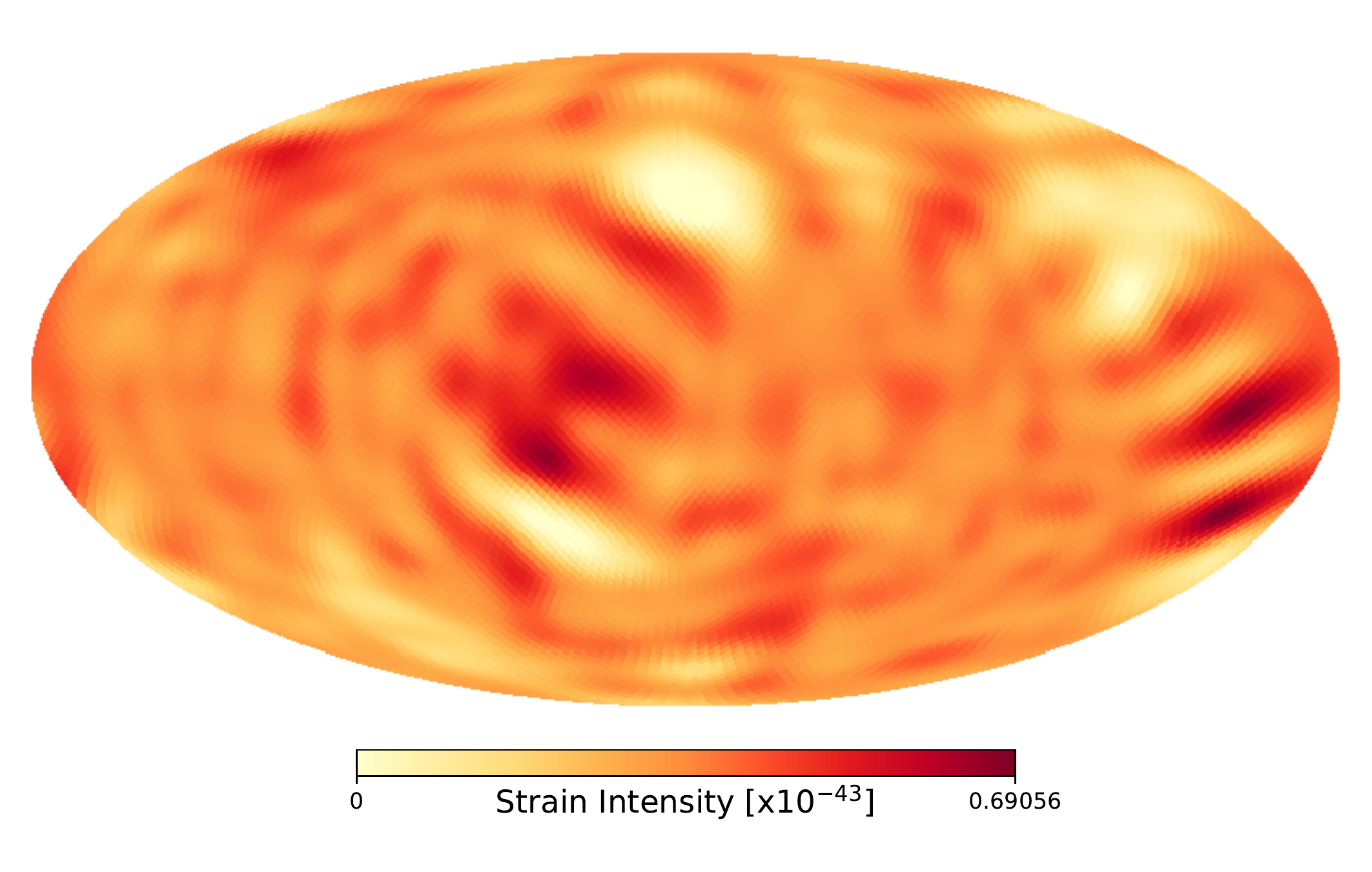}
\end{minipage}%
\begin{minipage}{.32\textwidth}
  \centering
  \includegraphics[width=0.95\linewidth]{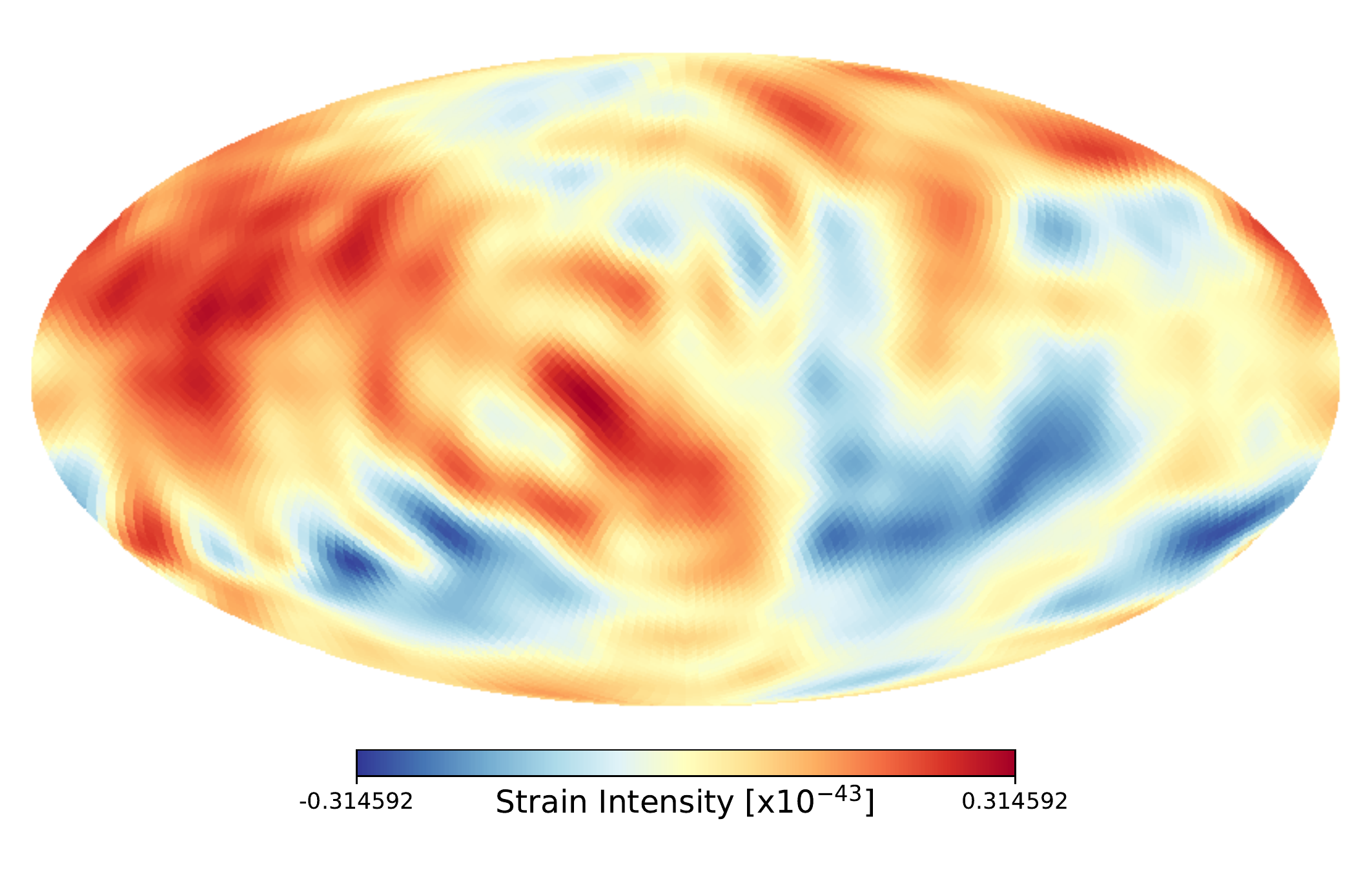}
\end{minipage}
\begin{minipage}{.32\textwidth}
  \centering
  \includegraphics[width=0.95\linewidth]{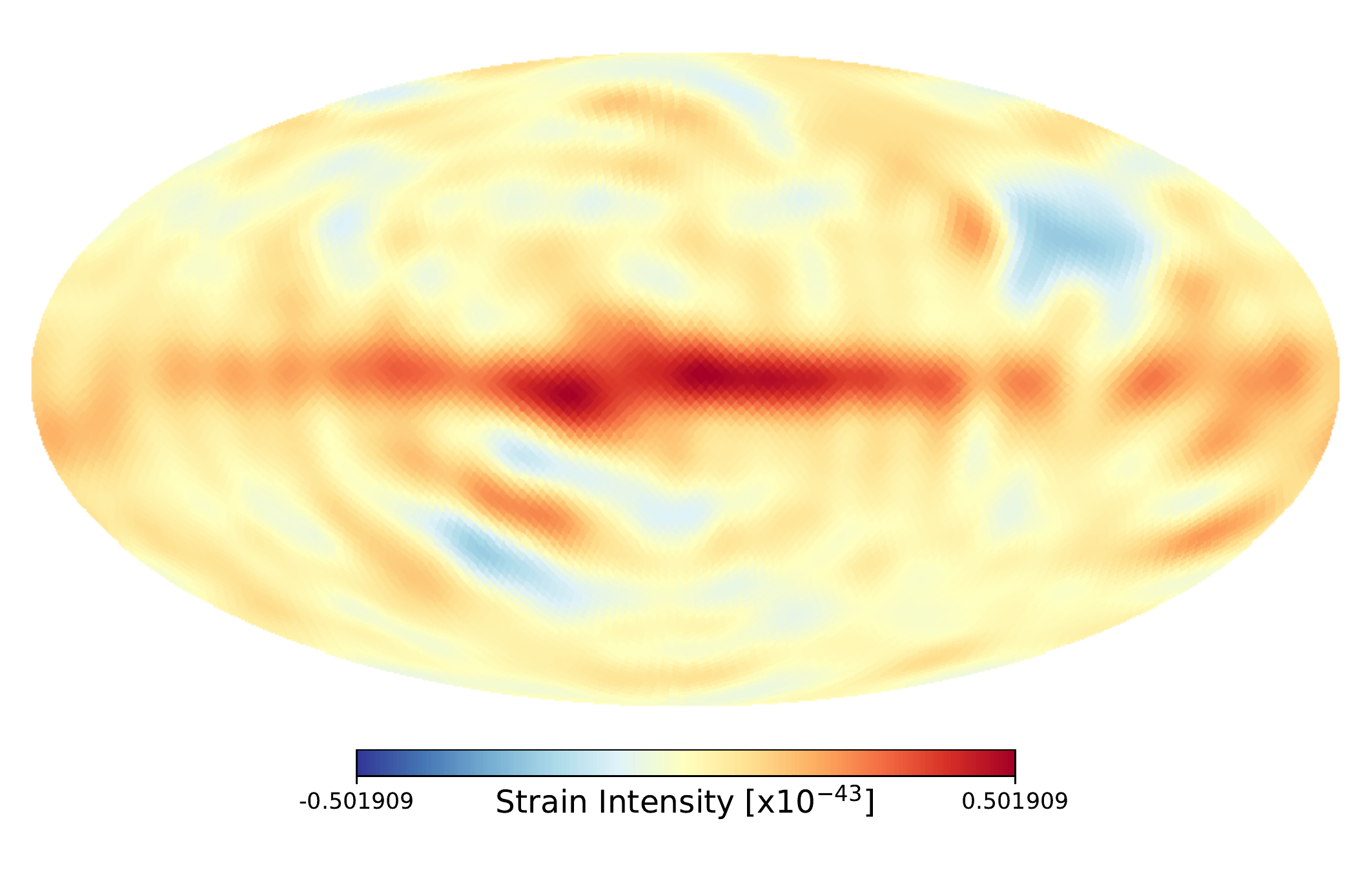}
\end{minipage}
\caption{Input ({\sl top}) and output ({\sl bottom}) maps for the low SNR cases for the same baseline combination as in Figure~\ref{fig:hsnr}. The signal level in these simulation was chosen so that the SNR is close to unity for the integration time.}
\label{fig:lsnr}
\end{figure*}

\subsection{Application to LIGO--type detector baselines}\label{sec:app_baselines}

In order to test our procedure in the most realistic setting from the outset we have based our algorithm around the LIGO open data format \citep{TheLIGOScientific:2016agk}. We load, sequentially, the LIGO ``O1'' data frames, stored with a sampling rate of 4096 Hz, where both LH and LL detectors were simultaneously in operation. When simulating data from more detectors we assume that they are operational for the same data frames. 

\begin{table}
\begin{center}
\caption{Frequency location $f$ (Hz) and widths $\sigma$ (Hz) for the inverse Gaussian notching applied to the LL and LH frequency data. The table covers the full frequency range of the data but only the notches highlighted with $^\dagger$ affect the band-passed data used to reconstruct the maps}
\label{tab:notch}
\begin{tabular}{l l || l l}
\hline
\hline
$f$ & $\sigma$& $f$ & $\sigma$\\
\hline
34.70& 0.5&120.00$^\dagger$ & 1.0\\
35.30& 0.5&179.99$^\dagger$ & 1.0\\
35.90& 0.5&304.99$^\dagger$ & 1.0\\
36.70& 0.5&331.90 & 1.0\\
40.95& 0.5&510.02 & 5.0\\
60.00& 0.5&1009.99 & 1.0\\
\hline
\hline
\end{tabular}
\end{center}
\end{table}


The data is parsed using the LIGO flag system in order to reject frames where injection events were present, periods when the detectors were being tested, or sections of the data were the LIGO pipeline has verified that one or either of the instruments was operating outside of the established parameter range. We then subdivide the remaining data into segments of 60 seconds. These constitute the time frames labelled as $\tau$ where, for the angular scales being targeted here, we can consider the motion of the Earth to be negligible. Each segment $\tau$ is edge--tapered with a cosine window over the first and last three seconds before Fourier transforming to the frequency domain. The frequency data is then notch filtered to remove biases due to known harmonics (see Table~\ref{tab:notch} for details) and band--passed to the range $f\in [80.0,\,300.0]$ Hz. This procedure is applied to all detector time streams involved in the number of baselines being considered.

A power spectrum is also computed from the filtered data and a three parameter model
\begin{equation}
P(f) = {\cal A}\left[\left(\frac{{\cal B}}{0.1+f}\right)^4+\left(\frac{0.04f}{{\cal C} }\right)^2+(0.07)^2\right]\,,
\end{equation} 
is is fitted to the data spectra, within the band--pass, to obtain a functional form of the noise spectra of each detector time stream. To avoid any residual bias due to harmonic lines the data is gapped around known harmonics when fitting with a width of $5\sigma$ either side of the notch frequency where the $\sigma$ correspond to those in Table~\ref{tab:notch}. Since LIGO data is noise dominated we do not expect any signal bias in this noise estimation procedure\footnote{At much higher SNR such as that expected for LISA this step will require an iterative procedure to obtain a noise spectrum that is not biased by any signal contribution.}. If the fit fails, or the parameters are outside a given range, the time segment is discarded, this happens for $\sim 3$\% of the remaining data. The failure typically happens for time segments where the $1/f$ tail is larger than expected or if there is an unexpected line present in the spectra. We use the product of the fitted models to obtain the weighting function used in the maximum-likelihood solution
\begin{equation}
P^{\tau, b}_f = \left[P_f^{\tau, A}P_f^{\tau, B}\right]^{1/2}\,,
\end{equation}
where $A$ and $B$ denote the two detectors defining the baseline labeled $b$. The cross--correlation of the two time streams
\begin{equation}
d_f^{\tau,b} = \langle s_f^{\tau,A} s_f^{\tau B\,^\star}\rangle\,,
\end{equation}
is then the ``observed'' data used in the mapping procedure. Figure~\ref{ligo_psd} shows the spectral density for a random segment $\tau$ of LL data along with the notching centres and resulting fit. The data is band--passed in order to exclude the large $1/f$ and $f$--noise tails. 

The map $\bm z$ and operator $\bm M$ are accumulated over each time segment $\tau$ and baseline $b$ in the combination being considered. We assume a spectral dependence $E_f={\rm const}$ in all the cases considered here. 

The algorithm is parallelised using MPI to distribute individual time segments $\tau$ between processors. Since the time segments are of equal length, the parallelisation is well balanced and scales close to linearly with number of MPI processes. Once all the time segments have been analysed we use (\ref{eq:ml_map}) to obtain the final maximum--likelihood map $\bm \tilde s$. 

To interpret the signal-to-noise of features in the map we require the pixel noise covariance. This is given by the inverse of the operator $\bm M$. In practice however, the operator $\bm A$ here introduces a relative calibration between the signal and noise in (\ref{eq:observe}) due to the rescaling of the windowed time streams. This calibration cancels out in the maximum--likelihood solution (\ref{eq:soln}) but needs to be accounted for in calculating the noise covariance of the map \citep{Bond:2001bd}. To calculate the scaling factor we evaluate the integral of the {\sl unweighted} operator $\sum_{f,\,\tau,\,b}A_{f\,p}^{\tau,\,b}$ and use this to renormalise the noise matrix. 
For visual comparison we also use the map of noise standard deviation
\begin{equation}
\sigma_p = \left[ M_{pp}^{-1}\right]^{1/2}\,,
\end{equation}
although we note that the correlations in the final maps are non-trivial and $\sigma_p$ is only a guide for the interpretation. 

We use a Healpix resolution level of $N_{\rm side}=8$ giving $N_{\rm pix}=768$ for the maximum--likelihood maps. This corresponds to a pixelisation scale $\sim 7$ degrees or a Nyquist scale of $\ell\sim32$. For the purpose of visualisation we smooth the resulting maps with a 10 degree Gaussian beam and then over--resolve to $N_{\rm side}=32$. 

\subsection{Simulations}

\begin{figure}
\centering
 \includegraphics[width=\columnwidth]{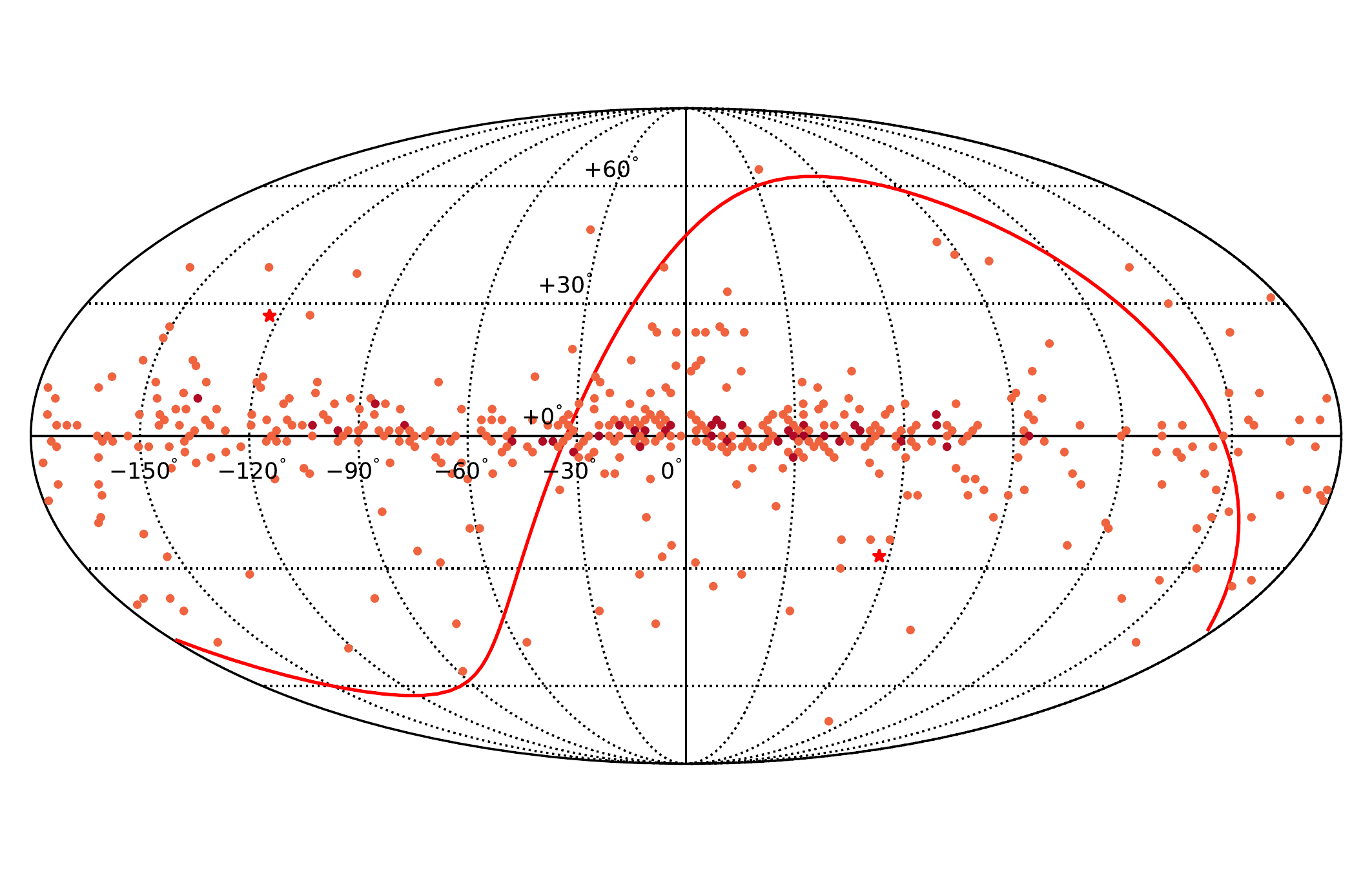}
 \includegraphics[width=\columnwidth]{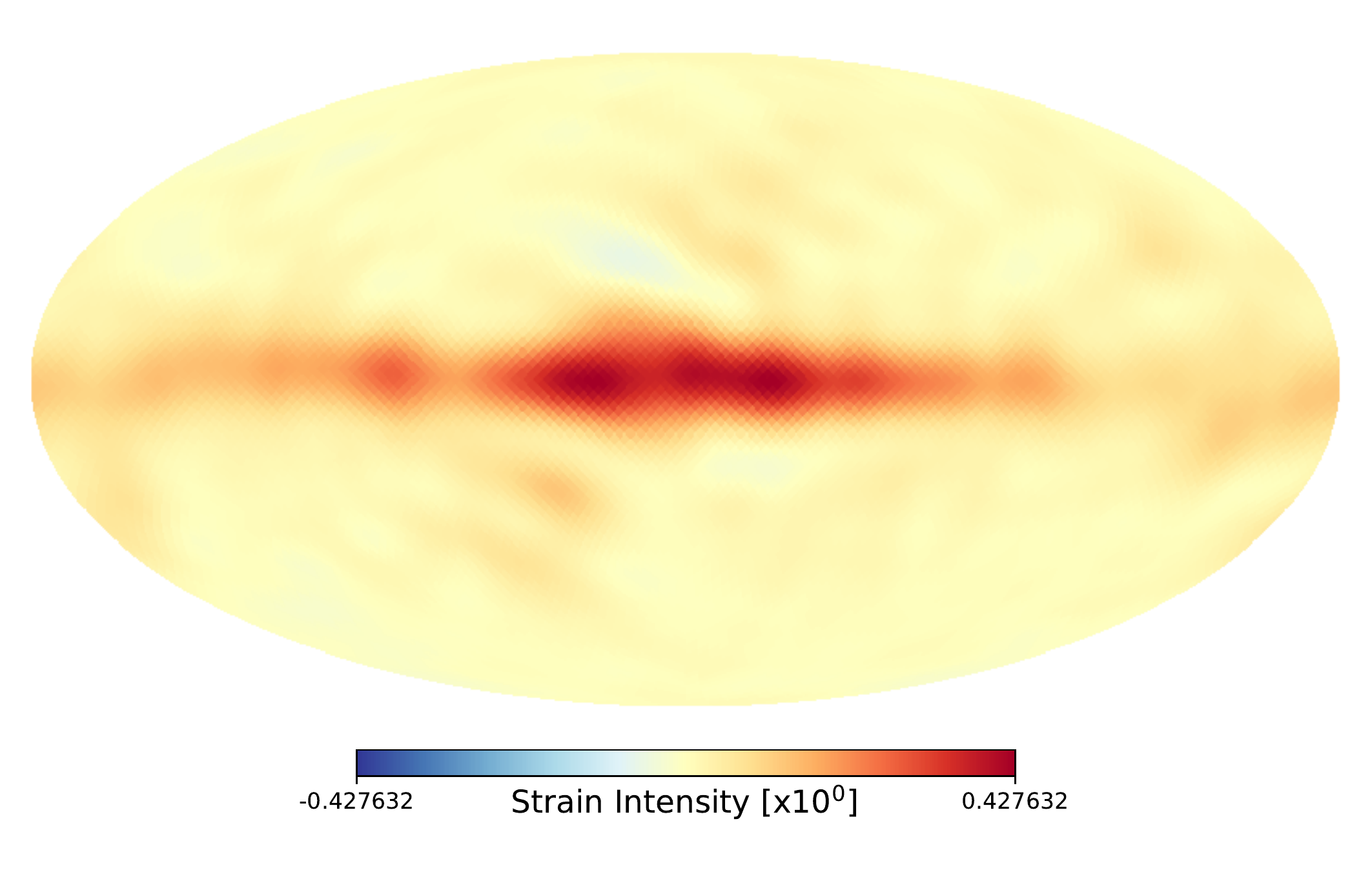}
\caption{The high SNR non--stationary ``Galaxy'' input case for approximately 36 hours of analysed mock LIGO-VIRGO data. The accumulated input Poisson map ({\sl top}). Each ``event'' is normalised to unit amplitude and the colour scale indicates the number of ``events'' accumulated in each pixel (maximum four in this case). The recovered map ({\sl bottom}). The increased integration time compared to the other tests is required to accumulate the signal of enough ``events''. After a sufficient number of ``events'' the map converges to the stationary case output map seen in Figure~\ref{fig:hsnr}.}
\label{fig:hpoi}
\end{figure}

\begin{figure}
\centering
\includegraphics[width=\columnwidth]{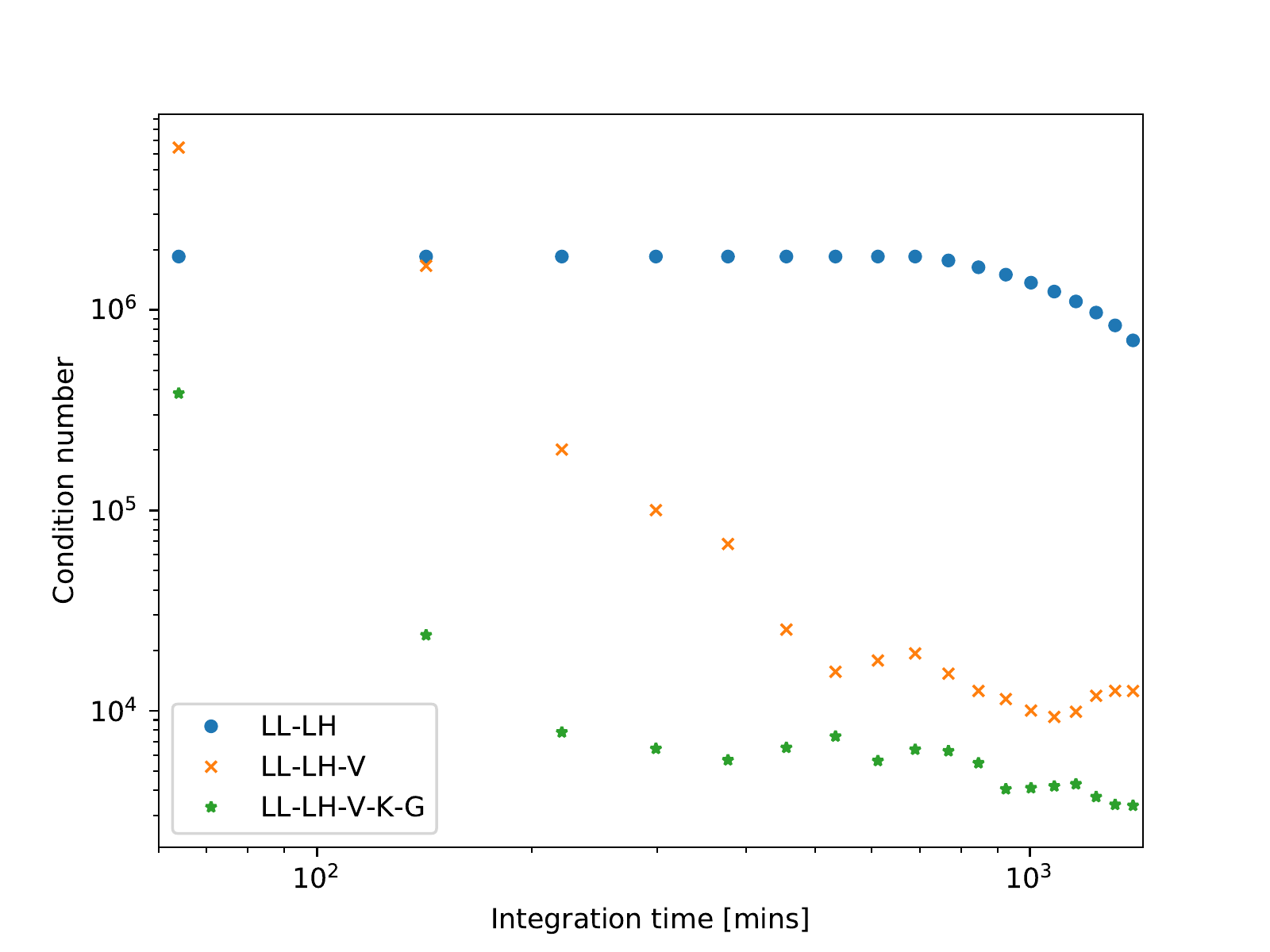}
\caption{Evolution of the condition number of the operator $\bm M$ with integration time for a number of different simulated baseline combinations (single, three, and ten baselines). The conditioning of the maximum-likelihood problem improves rapidly over the first 12 hours except for the one baseline (LIGO--only) case.}
\label{fig:cond_sim}
\end{figure}

For the purpose of testing our map--making procedure we we substitute the frequency domain data, after all filtering steps, with a scan of an input intensity map ${\bm I}^{\rm sim}$. This is obtained using the transposition of the projection operator
\begin{equation}
\begin{split}
	d_f^{\tau,b} &:= \sum_p A^{\tau,b}_{fp}\, I^{\rm sim}_p\\
    &= \Delta_f E^{}_f \frac{4\pi}{N_{\rm pix}}\sum_p \,\gamma^{\tau,b}_p\,I^{\rm sim}_p\, e^{i2\pi f\, \bm{b}^\tau\cdot \bm{\hat p}}\,.
\end{split}
\end{equation}
This procedure could be extended to obtain fiducial simulations of the time domain data by transforming back to a time stream in order to add transient systematics. Since we work in the frequency domain where the noise is also assumed to be uncorrelated this is not motivated at this stage. Noise is included by adding an uncorrelated, Gaussian realisation with variance, at each frequency, given by the fitted cross-correlation power spectrum $P_f$. The final product, at each 60 second time segment $\tau$, is a set of simulated signal {\sl plus} noise frequency data for each baseline. All baselines will have the same noise variance given by the actual LIGO data noise spectra for each time segment and gaps in the simulations will all mimic those resulting from missing of flagged LIGO data. 

To avoid any aliasing in the simulation procedure the input map ${\bm I}^{\rm sim}$ is generated at a resolution level $N_{\rm side}=32$ such that the structure in the input is over--resolved with respect to the working resolution of the output maps which is limited by the smoothing scale of most baselines (see Figure~\ref{fig:uvw}).

\section{Results}\label{sec:res}

\begin{figure}
\includegraphics[width=\columnwidth]{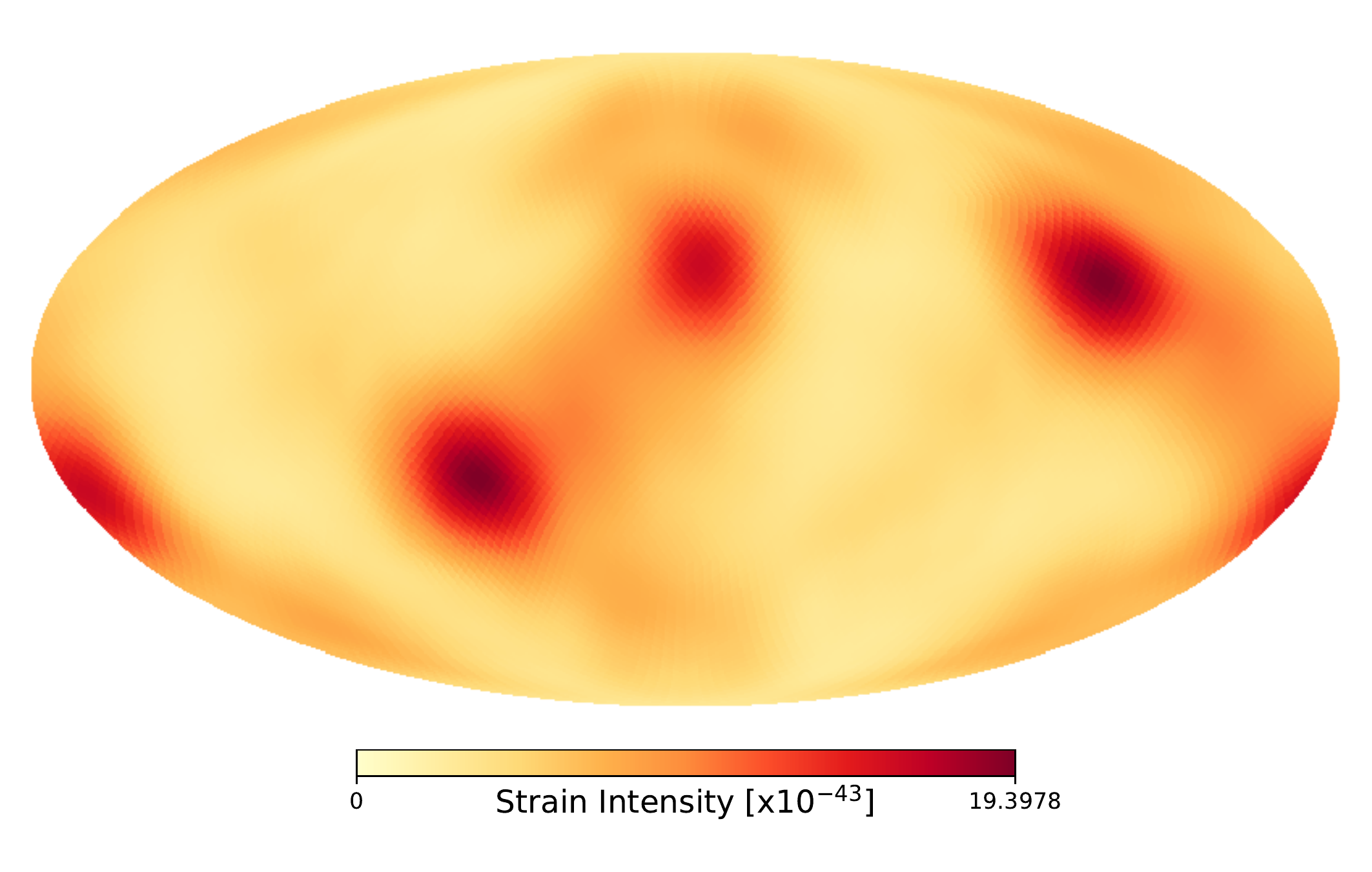}
\caption{Noise map for the ten baseline case after 24 hour of integration. The noise is inhomogeneous and there are significant pixel--pixel correlations that are not visualised in the map.}
\label{fig:5d_Mpp}
\end{figure}

A number of different choices for input maps are used to test the reconstruction with three separate Signal--to--Noise Ratio (SNR) levels:
\begin{itemize}
\item ``High'' SNR, corresponding to a noiseless case with input strain intensity of $h^2\sim {\cal O}(1)$.
\item ``Medium'' SNR, corresponding to a signal level comparable to the typical noise level of the LIGO baseline. This is chosen such that the maximum--likelihood maps converge after analysis of ${\cal O}(1)$ day data. 
\item ``Low'' SNR,  for this input map the signal level is chosen such that the estimated maps are still converging after on similar time scales.
\end{itemize}
Note that the high SNR does {\sl not} bias the noise estimation step described above as the noise is estimated using the actual LIGO data for each segment before it is substituted with the simulated scan. 

For each of the SNR cases we also consider different types of input maps. The first are pure ``monopole'' maps.  The second case  are statistically isotropic ``Gaussian'' random realisations of an $\ell(\ell +1)C_\ell={\rm const}$ angular power spectrum for $\ell \le 8$.  The the third case are anisotropic ``Galaxy'' maps obtained by smoothing the dust component Planck satellite map\footnote{\,https://pla.esac.esa.int} \citep{Adam:2015wua}. These are dominated by galactic dust emission and provide an input map where the signal is concentrated along the Galactic equator. These are not supposed to simulate any realistic GWB but are instead designed to test the algorithm with various SNR regimes and levels of anisotropy.

\begin{figure}
\centering
\includegraphics[width=\columnwidth]{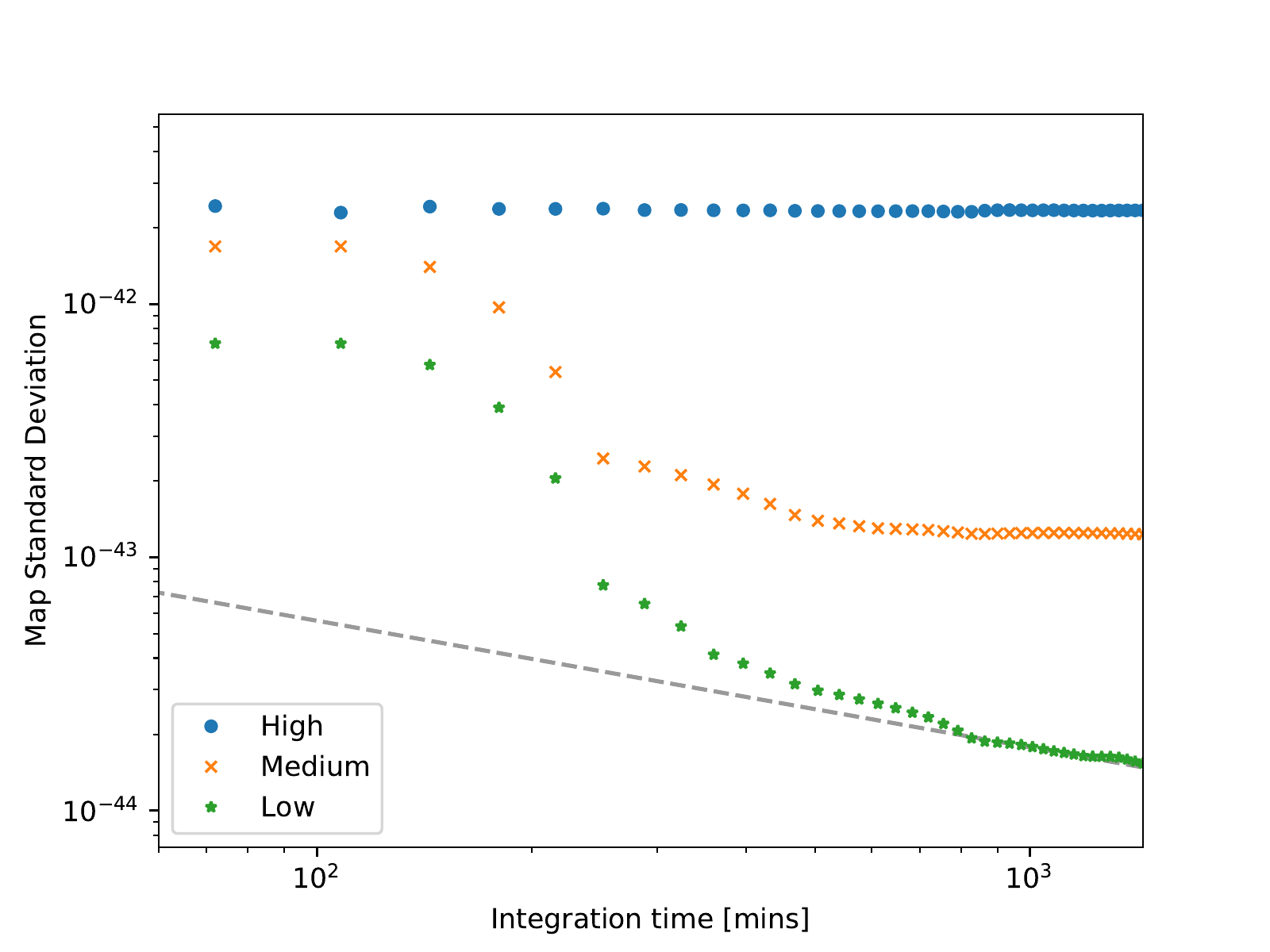}
\caption{ The map standard deviation for the LIGO-VIRGO, three baseline simulations with different SNR. The scaling of the standard deviation is initially affected by the ill--conditioning of the problem but enters the expected $t^{-1/2}$ scaling once the  rotation of the Earth has improved the conditioning. When the standard deviation becomes constant the maps have converged to a detection of the input signal. The only case that has not converged after a full Earth rotation is the low SNR one.}
\label{fig:stdev_sim}
\end{figure}

All the maps described so far are stationary in the sense that the strain intensity remains constant in time. We also run tests on a non-stationary case where we simulate a Poisson process at each time segment $\tau$ in each pixel. This generates input signals of a fixed amplitude according to a Poisson random draw with the mean of the Poisson distribution set to the ``Galaxy'' input map. The aim of this particular test is to show how the signal of a stochastic, non-stationary background, that may be arriving from distinct directions on the sky {\sl simultaneously}, is accumulated in the map--making procedure.  

Figure~\ref{fig:hsnr} shows the results obtained for a number of high SNR input maps after the integration of approximately 24 hours of mocked LIGO--VIRGO data made up of baseline formed by LIGO Hanford and Livingston detectors, and Virgo, as set out in Table~\ref{tab:dects}. A full Earth rotation is sufficient to reconstruct all $\ell \le 8$ modes included in the statistically isotropic cases accurately, although the effect of gaps in the LIGO data and the limited ``$uvw$'' coverage of the simulated LIGO--VIRGO baseline combination is seen in the induced pixel correlation that is most evident in the ``monopole'' case. The limited resolution of the observations, dictated by the angular scales of the overlap functions, is seen most clearly in the ``Galaxy'' case where the higher resolution input features are significantly smoothed by the convolution.

The low SNR,``Gaussian'' case, integrated over 24 hours, is shown in Figure~\ref{fig:lsnr}. The signal level for this test was set so that the residual noise in the maximum--likelihood map is comparable to the signal.

\begin{figure}
\begin{minipage}{\columnwidth}
\centering
\includegraphics[width=\columnwidth]{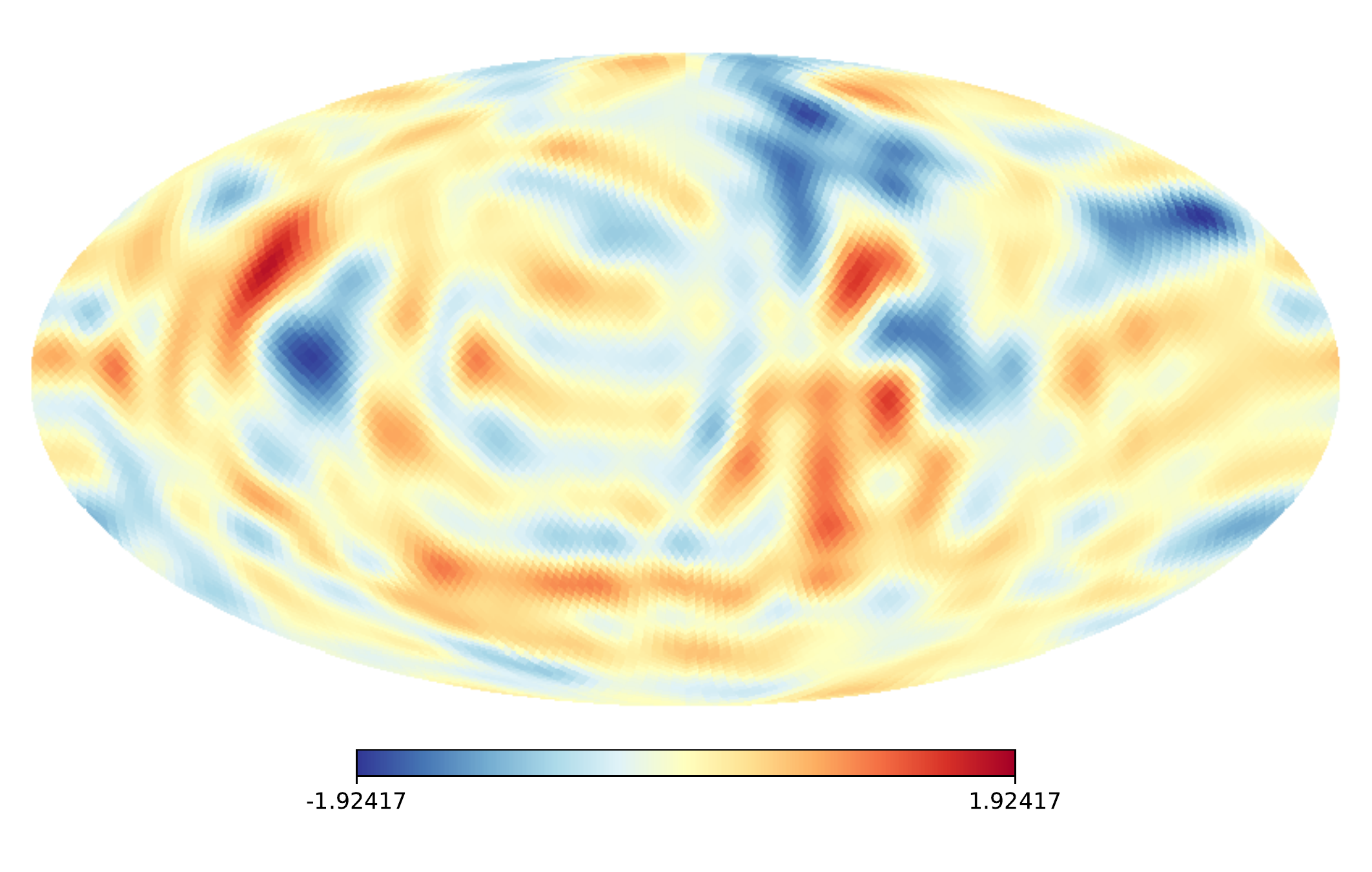}
\end{minipage}
\begin{minipage}{\columnwidth}
\includegraphics[width=\columnwidth]{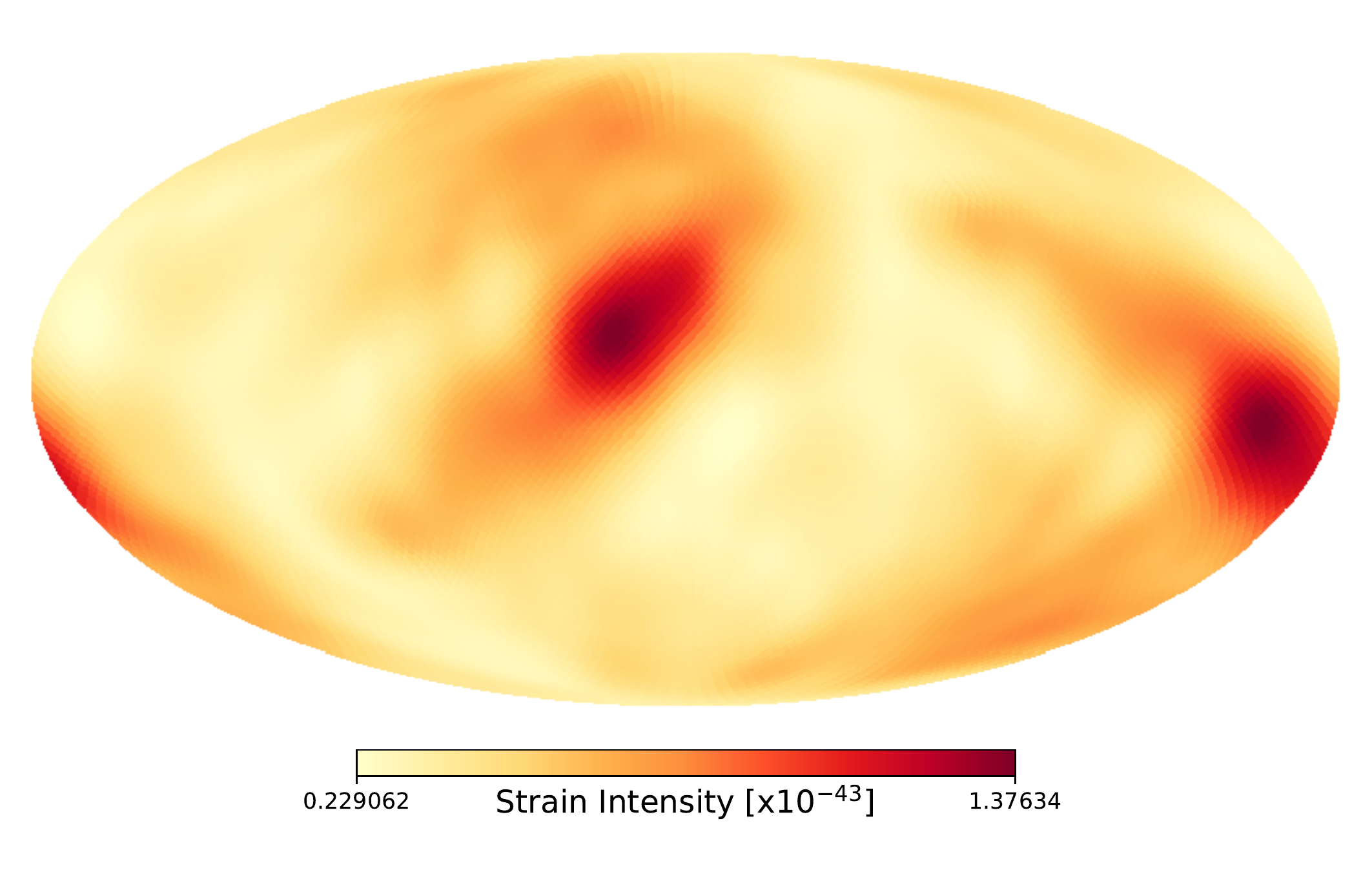}
\end{minipage}
\caption {Signal-to-noise ratio map for the integration of the first seven days LIGO ``O1'' release data ({\sl top}). The noise map ({\sl bottom}) for the run. A calculation using the full noise pixel covariance matrix gives a PTE of 32\% for the map.}
\label{fig:LIGO_maps}
\end{figure}

We estimate that, for current generation computing cores at a sampling rate of 4096 Hz, our algorithm can analyse 0.033 days of baseline data per core day with the cost scaling linearly in $\tau$ and $b$. To understand the convergence of the map estimation we look at the condition number of the operator $\bm M$ in (\ref{eq:ml_map}) and the standard deviation of the output maps as a function of amount of data analysed (in minutes). A large condition number, typically greater than ${\cal O}(10^8)$, is an indication that we are trying to reconstruct too many modes for the given observations - the problem is ill--conditioned. The standard deviation of the output maps should decrease approximately as the square root of the amount of time analysed as the noise is integrated down. A convergence of the standard deviation of the map to a constant value indicates that the map has converged to a given signal level which does {\sl not} average down with additional integration in time. 

Figure~\ref{fig:cond_sim} shows the condition value of $\bm M$ for the single, three, and ten baseline (see Section~\ref{sec:GWB}) case run on the high SNR ``Gaussian'' input case. The condition number is initially large for all cases but decreases rapidly over the course of 24 hours of integration time except for the single baseline case. In practice, the inversion in (\ref{eq:ml_map}) is conditioned using a pseudo inverse calculation with a limiting condition number of $10^5$. Thus after 24 hours, or a complete rotation of the Earth--based baselines, only the single LL-LH baseline case requires the removal of singular modes. This is an indication that for the single baseline case our working $N_{\rm side}=8$ resolution is sufficient and saturates the number of modes that can be reconstructed. The addition of more baselines increases the number of modes that can be reconstructed and the fact that we can carry out the inversion without removing singular modes is an indication that this resolution does not saturate the number of modes and could be increased further.

Figure~\ref{fig:stdev_sim} shows the evolution of the standard deviation of the maps as a function of integration time for various SNR regimes and cases. It is clear that, for the first few hours, the convergence is affected by the ill--conditioning of the problem.  However, after a approximately 6 hours of integration time, the standard deviation enters a scaling regime that is approximately close to the $t^{-1/2}$ limit expected until, for the higher SNR cases, it reaches a constant value indicating that the map has converged. 

\begin{figure}
\includegraphics[width=\columnwidth]{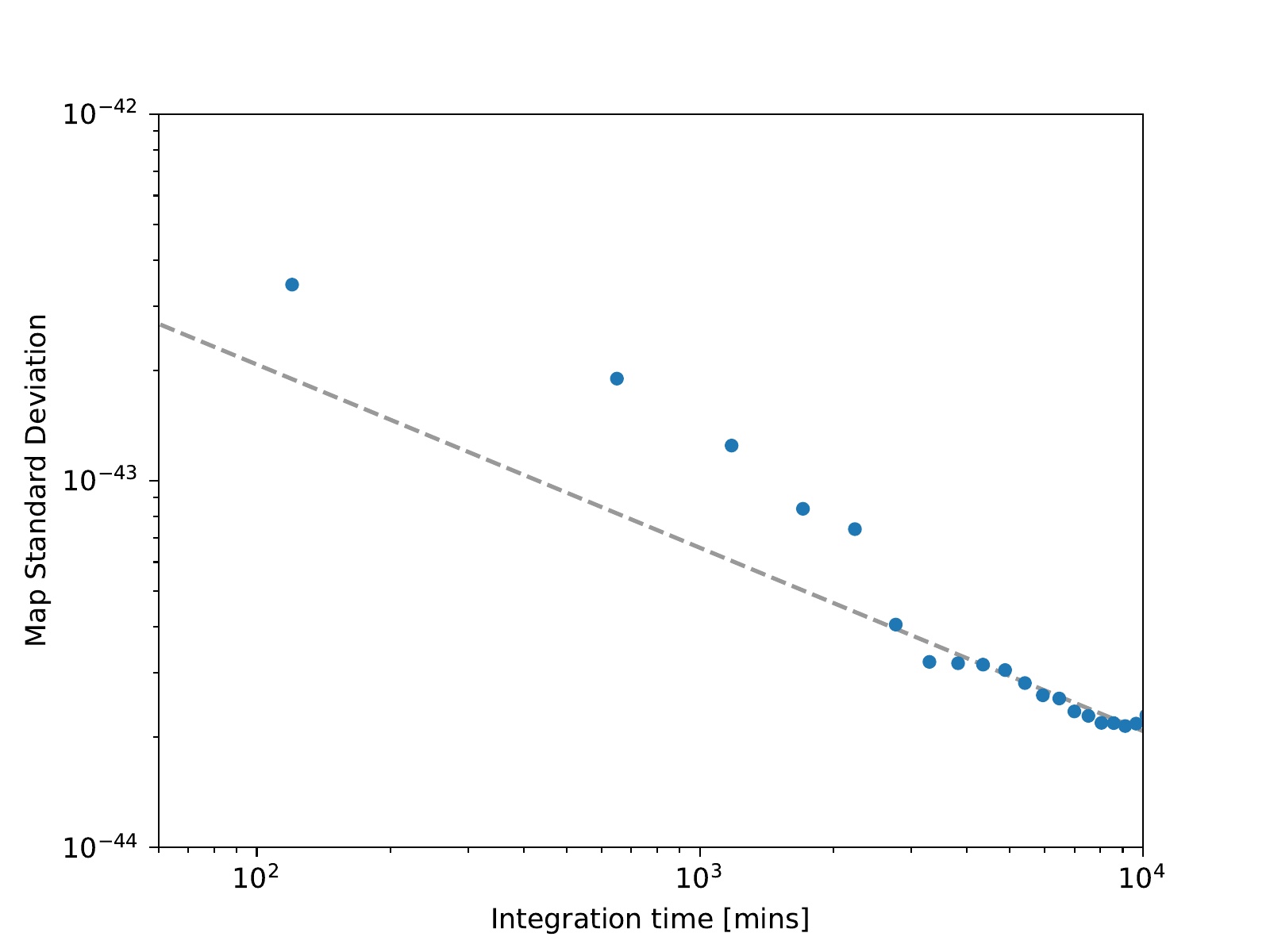}
\caption{The map standard deviation for the LIGO baseline. We observe that the map has not converged and is scaling as $t^{-1/2}$.}
\label{fig:LIGO_stdev}
\end{figure}

\subsection{Application to LIGO data}

In \cite{InPrep} we will present a complete analysis of the LIGO ``O1'' data release. Here we limit the application to the first seven days of data in order to test the algorithm on actual, albeit, noise dominated data. We carry out the same filtering, flagging, and estimation procedure on the data as described above. The conditioning of the reconstruction improves after a few days of integration as the effects of gaps in the data are minimised. 

Figure~\ref{fig:LIGO_maps} shows the SNR map for the integration obtained by dividing the maximum--likelihood map by the square root of the diagonal of the noise covariance matrix. We note again that the significant correlation in the noise together with the smoothing applied for visualisation purposes makes a visual interpretation of the significance of any structure in the maps difficult. In order to quantify the significance we calculate the actual $\chi^2$ of the maximum--likelihood using the full noise covariance matrix at the original $N_{\rm side}=8$ resolution level. The Probability To Exceed (PTE) statistic we obtain from this is 32\%. Thus all the structure in the map is consistent with a random realisation of the noise covariance, as expected given the LIGO noise amplitude compared to any potential GWB.

\section{Discussion}\label{sec:disc}

We have described the implementation and testing of a map--making method used to reconstruct the strain intensity on the sky through an incoherent integration of the cross--correlation of gravitational wave detector signals. Our method reconstructs the sky directly in the pixel basis and in galactic coordinates. The generalisation involved here is the implementation of all the transformations required to integrate the signal of a general, albeit Earth--based detector, directly onto the sky. As such this approach is not restricted to the detector frame as the definition of the coordinate system. 

The method has been built using a ``bottom--up'' approach by starting with existing LIGO data and modelling our generalised detector set on LIGO data formats and noise properties. This approach makes application to actual data as simple as possible as shown in the test analysis of a small subset of the most recently released LIGO data. In that case we obtain a map that is consistent with the pixel domain noise covariance. 

The simple input maps used to simulate the signal component in the test are not of much use in evaluating the potential for observations to detect either the monopole or anisotropic component of a stochastic background. A realistic simulation would require an injection of a properly simulated event rate distribution in the time stream of the various detectors. The first step would be to build a robust simulator of stochastic events from particular populations with the correct correlation on the sky. We will work to include these time domain simulations in our pipeline in a future implementation. This work will also require a robust simulation of the time domain noise which entails additional complications with respect to the frequency domain approach taken in this work.

An important focus for future development will be the adaptation to off--Earth detectors in order for the algorithm to be applied to the LISA data. This will require an integration with pointing and orientation data for LISA along with careful verification of the translation and generalisation of the coordinate transformations required for that case. The method, with suitable redefinitions of the coordinate transformations and observing operators, can also be applied to PTA data. These provide much longer baselines than those achievable by laser interferometers and could be the first to detect a GWB. 

In this work we have not considered the polarisation of the signal having focused solely on the estimate of the total intensity. Our treatment however, generalises easily to the reconstruction of all Stokes parameters, i.e. including $Q$, and $U$ linear polarisations, and the circular polarisation amplitude $V$. We will extend our algorithm to include these in future although we note that the polarisation of incoherent, stochastic backgrounds is not expected to be large due to the random orientation of each event contributing to the backgrounds. It would still be interesting to test this hypothesis of course and there may be some astrophysical information to be extracted from the polarisation. 

In \citet{InPrep} we will present a full analysis of the most recent, publicly released ``O1'' LIGO data set. The only limitation in such an analysis is computation time but having efficiently parallelised our method from the start we estimate that a full analysis of the available data ($\sim$ 5 months, single baseline) will take approximately one month with the resources currently available. Increasing the resolution of our maximum--likelihood maps further will come at a significant increase in computational cost since the method scales as $\sim N_{\rm pix}^2$. 

We do not expect the noise to integrate down to a level sufficient for a signal detection but the result of that work will provide an upper limit for $\Omega_{\rm GW}$ and any directional dependence and an independent verification of the LIGO collaboration limits in \citet{Abbott2017e}.

\section*{Acknowledgements}

AIR acknowledges support of an Imperial College Schr\"odinger Fellowship. This research was supported by STFC consolidated grants ST/L00044X/1 and ST/P000762/1.  We thank Laura Nuttall for useful discussions on LIGO data and its characteristics. We acknowledge Ciar\'an Conneely, Andrew Jaffe, Jo\~ao Magueijo, and Chiara Mingarelli for stimulating discussions. 


\bibliographystyle{mnras}
\bibliography{bib} 

\bsp	
\label{lastpage}
\end{document}